%% file: main.tex
\documentclass[]{JFM-FLM_Au}

\lefttitle{Murray B.C.~Kiernan et al.}
\righttitle{Resonant tides of Earth's core--BMO system}

\title{Resonant lunar tides of Earth's core and basal magma ocean}

\author{Murray B.C.~Kiernan\aff{1,2},
	Hamish C.F.C.~Hay$^{1,2}$,
	David W. Rees~Jones\aff{1}, 
        James F.J.~Bryson\aff{2}, \and Richard F.~Katz\aff{2}
}

\affiliation{\aff{1}School of Mathematics and Statistics, University of St Andrews, St Andrews, KY16~9SS, UK.
\aff{2}Department of Earth Sciences, University of Oxford, Oxford, OX1~3AN, UK.}

\corresau{Richard F.~Katz, richard.katz@earth.ox.ac.uk}

\input{newcommands} 

\begin{document}
\maketitle

\begin{abstract}
    Earth's magnetic field is generated by fluid motion in the liquid-metal core and has been active for billions of years. However, prior to the onset of inner-core growth, the sources of mechanical power that drove the geodynamo remain uncertain. During this period, the core may have been overlain by a basal magma ocean (BMO), creating two immiscible fluid layers separated by a density interface, beneath the solid mantle. We develop a theory for lunar tides in this core--BMO system, in which the tidal potential acts on the density contrast between the two layers. The resulting dynamics differ fundamentally from previous models driven by imposed deformation of the core's bounding envelope. In the inviscid limit, the response transitions from an equilibrium tide to an interfacial gravity wave as forcing frequency increases. The two regimes are separated by a resonance that occurs when the forcing frequency matches the natural frequency of the degree-2, order-2 interfacial mode. The inviscid core flow is formally identical to the canonical elliptical vortex, linking the problem to the theory of elliptical instability. Finite viscosity regularises the resonance, introduces phase lag and generates oscillatory boundary layers. Combined with parameterised models of lunar recession and BMO crystallisation, the theory predicts enhanced core-boundary ellipticity, core-flow speed, magnetic Reynolds number and instability metrics, particularly near resonance. These results identify a previously unexplored mechanism for tidally driven flow in differentiated planetary bodies and show that a BMO can enhance tidal coupling to the core, potentially contributing to dynamo action.
\end{abstract}

\section{Introduction}

Earth's core hosts the dynamo action that creates the terrestrial magnetic field. This field has seemingly been generated for at least the last 3.4~billion years, so has potentially shielded Earth from harmful solar radiation for most of its history \citep{tarduno2015hadean, weiss2015pervasive}.  Throughout this time, Earth has been subject to periodic gravitational forcing by the Moon. In models, lunar tides of the core are typically imposed via a prescribed, equilibrium displacement of the overlying mantle. More specifically, the core--mantle boundary is assumed to be an ellipsoid aligned with the Moon, and hence rotating differentially to Earth; the differential rotation gives rise to a tidal flow \citep{kerswell1994tidal, cebron2010systematic, landeau2022sustaining}. This formulation isolates the fluid dynamics and has led to a well-developed theory of tidal currents, elliptical instability and turbulent dissipation in planetary cores \citep[e.g.,][]{legal2015flows, vidal2026did}. The problem is relevant to the origin of Earth's magnetic field because lunar tides have been proposed as a potential source of mechanical power for driving the early geodynamo \citep{landeau2022sustaining}. However, the imposed-ellipticity formulation assumes that the fluid core is bounded directly by a solid mantle. This assumption is questionable for the early Earth, when a long-lived basal magma ocean (BMO) separated the liquid-metal core from the overlying solid mantle \citep{labrosse2007crystallizing, boukare2025solidification, lark2026coupled}.

The presence of a BMO in the early Earth changes the tidal formulation in a fundamental way. The region beneath the solid mantle then comprises two immiscible liquid layers: a dense metallic core overlain by a less dense, more viscous silicate melt. In this case, the lunar tidal potential acts on the density contrast to deform the core--BMO interface dynamically, rather than acting solely via the motion of the solid-mantle boundary. This raises fluid-dynamical questions that are distinct from the classical imposed-ellipticity problem. In particular, how does a tidally forced, self-gravitating, two-layer body respond when the interface between the layers is free to move, and how does this response depend on forcing frequency, BMO thickness and viscosity?

These questions are motivated, in part, by uncertainty in the power source of the early geodynamo. At the present day, the geodynamo is driven largely by compositional convection resulting from the gradual outward solidification of the inner core. In the deep past, before the onset of inner-core nucleation at $1\pm0.5$~billion years ago, this mechanism could not have been active \citep{biggin2015palaeomagnetic, bono2019young}. Nonetheless, palaeomagnetic evidence indicates that Earth generated a magnetic field at that time \citep{biggin2011palaeomagnetism, tarduno2015hadean, weiss2015pervasive, nichols2024possible}. Several other mechanisms have been proposed to power the earlier geodynamo, but each faces constraints \citep{landeau2022sustaining}. For instance, thermal convection driven by secular core cooling is the classical explanation, but its viability depends on the thermal conductivity of the liquid core, and on the rate at which heat can be extracted through the mantle. If the core conductivity is high, as suggested by ab initio calculations \citep{davies2015constraints-b1c}, the adiabatic heat flux is large and a thermally convecting core must cool rapidly. Such rapid cooling is difficult to reconcile with a young inner core unless the early core contained a large excess of sensible heat before crystallisation began \citep{williams2018thermal, driscoll2023new}. In that case, however, the hot core would also tend to melt the base of the mantle, promoting the formation of a BMO. Experimental studies have instead argued for substantially lower core conductivities, which would allow thermal convection at lower rates of core heat loss and make a long-lived, thermally driven dynamo more compatible with late inner-core nucleation \citep{andrault25}. The early-core energy budget therefore remains sensitive to unresolved discrepancies between ab initio and experimental estimates of core conductivity, both of which are subject to significant uncertainties and possible systematic errors. Thermochemical alternatives, including the exsolution or precipitation of light-element-bearing phases such as MgO or SiO$_2$ can, in principle, provide buoyancy before inner-core growth, but require favourable core composition, equilibration conditions and thermal evolution \citep{orourke2016powering, hirose2017crystallization, badro2018magnesium}.

A BMO introduces both an additional difficulty and, potentially, an additional region of dynamo action. A BMO would release latent heat during protracted crystallisation \citep{labrosse2007crystallizing, lherm2024thermal} and concentrate radiogenic, heat-producing elements \citep{lark2026coupled}. Both effects would tend to reduce heat extraction from the core and suppress a thermally driven core dynamo. Conversely, if the silicate melt were sufficiently electrically conductive, convection within the BMO itself could have generated a magnetic field \citep{stixrude2020silicate-dynamo, dragulet2025electrical}. However, recent dynamo calculations suggest that despite the possibility of strong fields in thin-shell calculations, a BMO-hosted dynamo is unlikely under realistic electrical conductivity, thermal evolution and rotational constraints \citep{schaeffer2025energetically}. 

Forcing by the lunar tide injects mechanical energy directly into the fluid core, rather than relying solely on convective buoyancy. Existing estimates of tidal power and its implications for dynamo generation, however, have not considered the effect of a BMO. Instead, they assume that the lunar tide deforms the solid mantle, which directly imposes an ellipticity on the core's outer boundary \citep{legal2015flows, landeau2022sustaining}, and promotes dissipation in a viscous boundary layer \citep{vidal2026did}. Even independent of dynamo implications, the presence of a BMO fundamentally alters how lunar tides couple to the core because the forcing acts on a deformable density interface rather than solely through the motion of a solid boundary.

Here we test the hypothesis that a BMO can amplify tidally driven motion in the core by formulating and analysing a theory for the dynamics of a core--BMO system. We consider a two-layer model consisting of an inviscid or weakly viscous metallic core overlain by a viscous silicate BMO, both subject to lunar tidal forcing and self-gravity. The outer boundary of the BMO is held fixed, so that the response arises from the tidal gravitational potential rather than from an imposed solid-mantle ellipticity. We work in a reference frame rotating with Earth's spin, so that the Moon has a relative angular frequency in the rotating frame.  We linearise about a hydrostatic reference state and solve for the spherical harmonic degree-two, order-two forced response. In doing so, we neglect nonlinear advection of momentum and the Coriolis force. These approximations isolate the primary interfacial gravity response of the layered system and allow the role of BMO thickness and viscosity to be analysed explicitly. The resulting theory yields an analytical inviscid solution and semi-analytical viscous solutions that resolve the boundary layers and the dissipation associated with finite viscosity. While the model is linear and so does not directly describe the turbulent flow, we analyse the potential for a transition to complex turbulent flow by estimating critical parameters for instabilities of the primary flow. 

The model predicts three main behaviours. First, at low forcing frequency, the response approaches an equilibrium tide controlled by a balance between the tidal potential and self-gravity. Second, at higher frequency, the density jump supports an interfacial gravity tidal wave, with its crest phase-shifted relative to the Moon. These regimes are separated by a resonance at which the forcing frequency matches the natural frequency of the degree-two interfacial mode. Viscosity regularises this inviscid resonance, introduces a phase lag and produces oscillatory boundary layers at the core--BMO boundary and at the top of the BMO. The strength of the viscous modification is controlled primarily by a BMO-Reynolds number based on the BMO thickness, so that thin or highly viscous BMOs damp the resonant response. Nevertheless, over a broad range of plausible early-Earth parameters, the BMO enhances the ellipticity of the core--BMO boundary, the characteristic core-flow speed, the magnetic Reynolds number and metrics for secondary instabilities relative to the equilibrium-tide response.

The remainder of the paper is organised as follows. Section~\ref{sec:formulation} defines the domain, boundary conditions, tidal forcing and dimensionless control parameters. Section~\ref{sec:linearisation} derives the linearised equations and general solution. Section~\ref{sec:inviscid-solution} analyses the inviscid limit, including the equilibrium tide, interfacial resonance and relation to the canonical elliptical vortex. Section~\ref{sec:viscous-solutions} analyses the full viscous model, and quantifies viscous damping, phase lags and dissipation. Section~\ref{sec:instabilities} evaluates the potential for magnetic induction and secondary instabilities, including elliptical and boundary-layer instabilities. Section~\ref{sec:discussion} discusses limitations of the theory, then potential implications for the early-Earth geodynamo by analysing orbital and BMO evolution over Earth's history. The section also considers the role of imposed solid-mantle forcing, relaxing the assumption that the outer boundary of the BMO is fixed. Finally, it considers other potential applications of the present model. Concluding remarks are in section~\ref{sec:conclusion}.

\section{Model formulation}
\label{sec:formulation}

In this section we pose a mathematical theory for lunar tides of the core--BMO system. The development is divided into four subsections. Through these, we arrive at a dimensionless system of partial differential equations, boundary conditions, forcing function, and control parameters. We estimate values of these dimensionless parameters.

\subsection{Domain and boundary conditions}

\begin{figure}
    \centering
    \includegraphics[width=\linewidth]{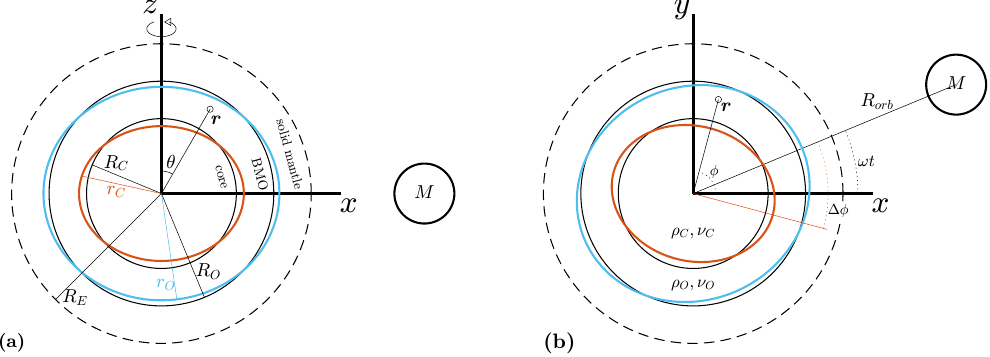}
    \caption{Schematic diagram of the domain (not to scale). The red line represents the outer boundary of the core; the blue line is the outer boundary of the BMO. \textbf{(a)} $x$--$z$ (meridional) plane. The $z$ axis is Earth's rotation axis and normal to the orbital plane. \textbf{(b)} $x$--$y$ (orbital and equatorial) plane. The blue BMO ellipse always has its long axis aligned with the Moon, but the long axis of the red CMB ellipse can have a non-zero phase lag $\Delta\phi$ relative to the Moon. Throughout the main text, we assume that the solid mantle is static, in which case the BMO outer boundary is undeformed; we consider its deformation in appendix~\ref{app:solid-mantle-forcing}.}
    \label{fig:domain}
\end{figure}

The domain, depicted in Figure~\ref{fig:domain}, comprises two concentric subdomains. The inner subdomain is Earth's liquid-metal core. In its spherical reference state, the core has a radius $R_C$. The silicate-liquid BMO surrounds the core and has a reference-state outer radius $R_O$. At radii beyond $R_O$ is the solid part of the mantle, which is outside of the domain.  A spherical coordinate system $(r,\theta,\phi)$ with position vector $\rpos$ has $\theta=0$ in the direction of Earth's rotation axis.  This axis defines the $z$ direction, and we assume that it is also a normal to the orbital plane of the Moon. The $x$ axis resides in this plane and pierces a fixed location on Earth's equator. Therefore, the coordinate system rotates anticlockwise around the $z$ axis over time $t$ with an angular frequency of Earth's spin, $\omega_\text{spin}$. In this rotating system, azimuth $\phi$ is measured anticlockwise from the $x$ axis. 

Under tidal deformation, the interfacial boundary between the core and BMO deviates from spherical. It is described by a surface of radius $r = r_C(\theta,\phi,t)$, with the denser core at $r < r_C$. We refer to $r_C$ as the core--magma boundary, which we abbreviate to CMB (in the zero-thickness limit of the basal magma ocean, the CMB is the core--mantle boundary). The surface $r_C$ is endowed with unit normal and tangent vectors $\nvec,\, \tvec^\theta,\, \tvec^\phi$. In the absence of tidally induced perturbations, $r_C = R_C$, a constant (because we neglect the centrifugal force).

The outer boundary of the BMO can also deviate from spherical, and this surface is denoted $r_O(\theta,\phi,t)$. Distinct from the surface $r_C$, which is a part of the solution, $r_O$ is specified in formulating the boundary conditions. Through the main body of the text, we take $r_O = R_O$ fixed, consistent with the solid having zero velocity in the rotating reference frame, $\vel_S=0$.  We relax this constraint and consider motion of the BMO outer boundary in appendix~\ref{app:solid-mantle-forcing}. 

The densities within subdomains $\rho_C>\rho_O$ and kinematic viscosities $\nu_C<\nu_O$ are distinct, uniform and given as problem parameters.  The tidally driven velocities $\vel_C(\rpos,t)$ and $\vel_O(\rpos,t)$ are fields to be determined as part of the solution. These velocities (and the associated viscous stresses $\stressten_C(\rpos,t)$ and $\stressten_O(\rpos,t)$) satisfy the following boundary conditions,
\begin{subequations}
    \label{eq:boundary-conditions-dimensional}
    \begin{align}
        \label{eq:bc-dimensional-rO}
        \left(\vel_O-\vel_S\right)\vert_{r_O} &= \boldsymbol{0}\quad\text{(no slip at BMO--solid-mantle boundary)},\\
        (\vel_C - \vel_O)\vert_{r_C} &= \boldsymbol{0}\quad\text{(continuity of velocity at core--BMO interface)},\\
        \nvec\cdot(\stressten_C - \stressten_O)\vert_{r_C} &= \boldsymbol{0} \quad \text{(continuity of stress at core--BMO interface)}, \\
        \partial_t{{r}_C} + \vel|_{r_C}\cdot\Grad{r}_C &=  \vel|_{r_C}\cdot\rhat \quad \text{(kinematic condition of core--BMO interface)},
    \end{align}
\end{subequations}
where the velocity of the solid mantle is defined as $\vel_S$. 

In addition to conditions \eqref{eq:boundary-conditions-dimensional}, we require that the total gravitational potential $\Psi$ is continuous and differentiable at the core--BMO interface. Perturbations to the gravitational potential are required to be 0 at $r=0$ and decay to 0 as $r\to\infty$.

\subsection{Governing equations}
In both fluids, the viscous-stress tensor (tension positive) is given by $\stressten = -P\identity + \mu\left[\Grad\vel + (\Grad\vel)^T\right],$ where $P$ is the full pressure and $\identity$ is the identity tensor. The dynamics are described by Navier-Stokes equations for incompressible flow, coupled with Poisson's equation for the self-gravitational potential,
\begin{subequations}
    \label{eq:governing-system-dimensional}
    \begin{align}
        \Div\vel &= 0,\\
        \label{eq:navier-stokes-dimensional}
        \pd{\vel}{t} + \vel\cdot\Grad\vel &= \nu\delsq\vel - \frac{\Grad P}{\rho} - \Grad(\Psi-\psi),\\
        \delsq\Psi &= 4\pi G \rho,
    \end{align}
\end{subequations}
where $\Psi$ is the planet's self-gravitational potential, $\psi$ is the tidal potential (with a positive sign convention) due to a moon and $G$ is the universal gravitational constant.  Since $\rho$ and $\nu$ take values specific to the two fluids, these equations must be solved separately in each of the subdomains, subject to the boundary conditions.

Although the governing equations are taken in a rotating reference frame, we neglect the Coriolis force from equation~\eqref{eq:navier-stokes-dimensional}. This approximation is difficult to justify \textit{a priori}, because Earth's rotation influences large-scale fluid motion and modifies the spectrum of free oscillations. However, omitting the Coriolis force simplifies the mathematics and allows the effects of self-gravity, viscosity and the density contrast at the core--BMO interface to be isolated. We therefore regard the present model as a limiting case that identifies the fundamental two-layer tidal response.  Preliminary calculations that retain rotation indicate that the dominant resonance described below persists and undergoes only a modest frequency shift. Additional resonances associated with rotationally modified interfacial gravity modes emerge at higher forcing frequencies through mode coupling, but the overall structure of a low-frequency, equilibrium-tide regime separated from a higher-frequency wave regime by a dominant resonance is preserved.  This behaviour is consistent with the broader literature on rotating tidal systems in which rotation modifies the modal structure without eliminating the primary tidal response \citep[e.g.,][]{rovira2023thin, Pontin2023b, Pontin2023a}. We return to the consequences of rotation, and to evidence supporting this interpretation, in the Discussion~(\S\ref{sec:discussion}).

\subsection{Tidal forcing}
\label{sec:tidal-forcing}

We assume that the Moon orbits in Earth's equatorial plane at a radius $\Rorb$, with an angular orbital frequency $n$. Earth spins with an angular frequency $\omega_\text{spin}$, which is approximately 27$\times$ larger than $n$ today, and even larger in the past \citep{farhat2022resonant} (both $\omega_\text{spin}$ and $n$ are positive, representing right-handed rotation about the $z$ axis).  In a reference frame that is rotating with Earth's spin (i.e., Fig.~\ref{fig:domain}), the apparent orbital frequency is $n-\omega_\text{spin} < 0$. However, to avoid defining the frequency $\omega$ as a negative quantity, we let 
\begin{equation}
    \omega \equiv \omega_\text{spin} - n \approx \omega_\text{spin} > 0
\end{equation}
and allow the lunar phase to progress as $\phi+\omega t$, rather than $\phi-\omega t$. Parameter values are listed in Table~\ref{tab:parameters}.

In the rotating frame and using the frequency $\omega$, we write the lunar tidal potential as
\begin{equation}
    \label{eq:tidal-potential-general}
    \tidepot(\rpos,t) = \frac{1}{2}\lmsumh \tidepot\elem r^\ell \,\sph{\ell}{m}(\theta,\phi + \omega t) + \cc
\end{equation}
where $\tidepot\elem$ is the amplitude of the degree-$\ell$ and order-$m$ spherical harmonic $\sph{\ell}{m}(\theta,\phi+\omega t)$. Because we assume a circular orbit with constant Earth--Moon separation, the $m=0$ components are independent of time and, hence, are neglected from the tidal potential. We use $\cc$ to represent the complex conjugate of the preceding expression.

Here we use non-normalised spherical harmonics, 
\begin{displaymath} 
    \sph{\ell}{m}(\theta,\phi+\omega t) = \e^{im(\phi+\omega t)}\legendre{\ell}{m}(\cos\theta),
\end{displaymath}
where $\legendre{\ell}{m}(\cos\theta)$ is the associated Legendre function of degree $\ell$ and order $m$ \citep{arfken1999MathematicalMethods}. 

Although we develop the analysis below for the general case of arbitrary $\ell,m$, we evaluate it for the simplest model of a lunar tidal potential, where the Moon makes a circular orbit in Earth's equatorial plane. This case considers only the $\ell=m=2$ mode for which 
\begin{equation}
    \label{eq:lunar-tidepot-magnitude}
    \tidepot_2^{2} = \tfrac{1}{4}GM/\Rorb^3,
\end{equation}
where $M$ is the mass of the Moon.  Note that $\sph{2}{2}(\theta,\phi+\omega t) = 3\e^{i2(\phi+\omega t)}\sin^2\theta$, which has an equatorial amplitude of 3.

The problem can also be forced by an imposed deformation of the solid mantle, via the boundary condition \eqref{eq:bc-dimensional-rO} at $r_O$. This approach is closely aligned with previous work on lunar tides of the core \citep[e.g.,][]{legal2015flows}.  In the main text, we focus on flows driven solely by the tidal potential, but consider non-zero $\vel_S$ in Appendix~\ref{app:solid-mantle-forcing}.

\subsection{Non-dimensionalisation and parameter estimates} \label{sec:non-dimensionalisation}

\begin{table}
    \centering
    \begin{tabular}{lcccll} 
         Name & symbol &\multicolumn{2}{c}{value or range}& units & note \\
          &  & today & early $\oplus$ &  &    \\ 
         \hline 
         Moon orbital radius & $\Rorb$ & $3.85$ & 0.38--3.8 & $10^5\,$km & [1]  \\
         Moon mass & $M$ & $7.3$ & same & $10^{22}\,$kg & \\
         Forcing frequency & $\omega$ & 7.0 & 7--70 & $10^{-5}\,$rad/s & See section~\ref{sec:tidal-forcing}. [1] \\
         Solid-mantle ellipticity & $\beta$ & $10^{-7}$ & & - & \\
         Core radius & $R_C$ & $3485$ & same & km & \\
         Core mass & $M_C$ & $181$ & same & $10^{22}\,$kg & \\
         Core density & $\rho_C$ & $10.0$ & same & $10^3\,$kg/m$^3$ & Present outer core density\\
         Core kin.~viscosity & $\nu_C$ & $0.1$--$1$ & same & $10^{-6}\,$m$^2$/s & dynamic viscosity $\sim$3 mPa~s [2] \\
         Core mag.~diffusivity & $\eta$ & 1 & same & m$^2$/s & $1/(\text{electrical cond.}\times\text{magnetic perm.})$ [3] \\
         BMO radius & $R_O$ & $0$ & $3485$--$5250$ & km & \\
         BMO density & $\rho_O$ & - & $5$ & $10^3\,$kg/m$^3$ & At bottom-mantle pressure [4] \\
         BMO kin.~viscosity & $\nu_O$ & - &  $2$--$20$ & $10^{-6}\,$m$^2$/s & dynamic viscosity $\sim$10--100 mPa~s [5] \\ \hline 
         
         Distance scale  & $[\rpos]$ & \multicolumn{2}{c}{$3.485\times10^6$} & m & $R_C$ \\
         Time scale & $[t]$ & \multicolumn{2}{c}{345} & s & $(4\pi G\rho_C)^{-1/2}$ \\
         Velocity scale  & $[v]$ & \multicolumn{2}{c}{$10^4$} & m/s & $[\rpos]/[t]$ \\
         Energy scale  & $[E]$ & \multicolumn{2}{c}{$4.3\times10^{31}$} & J & $4\pi G \rho_C^2 R_C^5$ \\
         Dissipation-rate scale  & $\left[\dot{E}\right]$  & \multicolumn{2}{c}{$1.2\times10^{29}$} & W & $[E]/[t]$ \\ \hline
         Reynolds number & $\re$ & \multicolumn{2}{c}{$3.5\times (10^{16}$--$10^{17})$} & - & $R_C^2\sqrt{4\pi G\rho_C}/\nu_C$ \\
         Density ratio & $\roc$ & \multicolumn{2}{c}{$1/2$} & - & $\rho_O/\rho_C$ \\
         Viscosity ratio & $\noc$ & \multicolumn{2}{c}{$20$} & - & $\nu_O/\nu_C$ \\
         Forcing frequency & $\Omega$ & \multicolumn{2}{c}{$0.025$--$0.1$} & - &  $\omega/\sqrt{4\pi G\rho_C}$ \\
         BMO outer radius & $\curlyr$ & \multicolumn{2}{c}{$1.001$--$1.2$} & - & $R_O/R_C$ \\
         Solid mantle ellipticity & $\mathcal{B}$ & \multicolumn{2}{c}{} & & $\beta/3\tidepotm$ \\
         Magnetic Prandtl number & $\pr$ & \multicolumn{2}{c}{$10^{-6}$} & - & $\nu_C/\eta$ \\
         Lunar tide pot.~magnitude & $\tidepotm$ & \multicolumn{2}{c}{$2.5\times(10^{-9}$--$10^{-6})$} & - & $(M/M_C)(R_C^3/\Rorb^3)/12$ 
     \end{tabular}
    \caption{\textbf{Parameters and their approximate values or ranges.} BMO is an abbreviation of basal magma ocean. Notes: [1] \cite{farhat2022resonant}; [2] \cite{li2021atomic, mineev2004viscosity, pozzo2013transport-564}; [3] \cite{davies2015constraints-b1c}; [4] \cite{caracas2024thermal}; [5] \cite{Sun2020, bajgain2022insights, huang2024low-4ff}. }
    \label{tab:parameters}
\end{table}

 We non-dimensionalise variables in \eqref{eq:boundary-conditions-dimensional} and \eqref{eq:governing-system-dimensional} with the  characteristic scales
\begin{equation}
    \label{eq:characteristic-scales}
    \begin{aligned}
    [\rpos] = R_C,\quad[t] = (4\pi G\rho_C)^{-1/2},\quad [\vel] = [\rpos]/[t],\quad  [\rho]=\rho_C,\\
    [\nu] = \nu_C, \quad \left[\Psi,\tidepot\right] = 4\pi G\rho_CR_C^2, \quad [P,\stressten] = 4\pi G\rho_C^2R_C^2.
    \end{aligned}    
\end{equation}
The characteristic time $[t]$ is the self-gravitational timescale. Using these scales to define dimensionless symbols with primes (e.g., $\vel = [\vel]\vel'$), substituting into the governing equations \eqref{eq:governing-system-dimensional}, and dropping primes gives
\begin{equation}
    \label{eq:governing-system-nondimensional}
    \Div\vel = 0,\qquad 
    \pd{\vel}{t} + \vel\cdot\Grad\vel = \frac{\nu}{\re}\delsq\vel - \frac{\Grad P}{\varrho} - \Grad(\Psi-\psi), \qquad 
    \delsq\Psi = \varrho,
\end{equation}
where all symbols are now dimensionless and we have defined
\begin{subequations}
    \begin{align}
    \label{eq:piecwise-viscosity}
        \nu &= \begin{cases}
            1 & r<r_C,\\
            \nu_O/\nu_C \equiv \noc & \text{otherwise},
        \end{cases}\\
        \label{eq:density-ratio}
        \varrho &= \begin{cases}
            1 & r<r_C,\\
            \rho_O/\rho_C \equiv \roc & \text{otherwise},
        \end{cases}\\
        \label{eq:reynolds-number-definition}
        \re &\equiv \frac{R_C^2(4\pi G\rho_C)^{1/2}}{\nu_C}.
    \end{align}
\end{subequations}
The last of these three is a Reynolds number for the core. The present definition uses a characteristic speed associated with the self-gravitational timescale.  This differs from the commonly assumed speed of flow in the core \citep{landeau2022sustaining}, and hence takes larger values than usual.  However we find below that under the present scaling, the velocity magnitude in our solutions is $O(1)$.

Three more dimensionless parameters arise in rescaling the domain size and the tidal potential,
\begin{subequations}
    \begin{align}
        \curlyr &\equiv R_O/R_C,\\
        \label{eq:Omega-definition}
        \Omega &\equiv  \omega\left(4\pi G\rho_C\right)^{-1/2},\\
        \label{eq:dimless-tidepot-definition}
        \tidepotm &\equiv \tidepot_2^2R_C^2/[\tidepot] = \frac{1}{12}\frac{M}{M_C}\left(\frac{R_C}{\Rorb}\right)^3.
    \end{align}
\end{subequations}
The first of these is the dimensionless radius of the top of the BMO (the thickness of the BMO is $\curlyr-1$); the second is the dimensionless forcing frequency. The third, $\tidepotm$, comes from equations~\eqref{eq:tidal-potential-general} and \eqref{eq:lunar-tidepot-magnitude} and represents the dimensionless magnitude of the $2,2$ lunar tidal potential at the CMB. 

Table~\ref{tab:parameters} summarises the dimensional parameters, the characteristic scales, and the dimensionless parameters, each with a value or range relevant for the Earth--Moon problem.  There are five main dimensionless parameters that exert a non-trivial control on the solution,  $\re,\,\roc,\,\noc,\,\Omega,$ and $\curlyr$. 
Amongst these main five dimensionless parameters, we consider the density ratio to be reasonably well constrained at $\roc=1/2$.  The Reynolds number is very large, and hence the core is nearly inviscid. The BMO is at least 20$\times$ more viscous than the core, and possibly more than that if it bears a slurry of crystals. The thickness of the BMO, given in non-dimensional terms by $\curlyr-1$, has varied over time from about $0.15$ shortly after the Moon-forming impact to (nearly) zero today \citep{schaeffer2025energetically}.  The dimensionless frequency of Earth's spin has decreased over this time period, corresponding to a range in $\Omega$ from $0.1$ for the early Earth to $0.025$ today. In analysing the tidal problem, we focus on the effects of $\Omega$ and $\curlyr$, and also consider the effect of $\noc$. 

There are three other non-dimensional parameters that are of lesser importance than the five discussed above. First, in the linearisation (\S\ref{sec:linearisation}), all the departures from the hydrostatic state are linearly proportional to the magnitude of the lunar tidal potential $\tidepotm$. Second, Appendix~\ref{app:solid-mantle-forcing} develops cases where the solid mantle deforms with imposed ellipticity $\beta$; this leads to an additional dimensionless parameter $\curlyb \equiv (\beta/3)/\tidepotm$ representing the amplitude ratio of forcing mechanisms. We take $\curlyb=0$ in the main text. Third, while not affecting the solution itself, the magnetic Prandtl number $\pr\equiv\nu_C/\eta$ indicates the potential for magnetic induction to overcome diffusion; $\eta$ is the magnetic diffusivity of the core.

\section{Finding a linearised solution}  
\label{sec:linearisation}

In this section we analyse the dimensionless problem of \S\ref{sec:non-dimensionalisation} by assuming that tides create a small perturbation to an otherwise hydrostatic state.  We obtain this reference state in the first subsection. Then, in the following three subsections, we linearise the equations, develop a general solution, and discuss how boundary conditions are used to determine solution constants (application of the boundary conditions is presented in detail in Appendix~\ref{app:BCs}),

\subsection{Hydrostatic state}

In the absence of a tidal forcing, the fluid attains hydrostatic equilibrium, denoted with a subscript $h$ and having $r_C = 1$. The hydrostatic pressure and total potential satisfy the system \eqref{eq:governing-system-nondimensional} with $\vel=\boldsymbol{0}$,
\begin{equation}
    \fd{}{r} P_h +\varrho_h\fd{}{r}\Psi_h = 0, \qquad 
        \fd{}{r} r^2\fd{}{r}\Psi_h = \varrho_h r^2.
\end{equation}
As boundary conditions, we impose continuity of $P_h$, $\Psi_h$, and $\infd \Psi_h/\infd r$ at $r=1$. The hydrostatic solution is
\begin{subequations}
    \label{eq:hydrostatic-solution}
    \begin{align}
        P_h(r) &= P^* - \varrho_h\Psi_h - (1-\varrho_h)/6,\\
        \Psi_h(r) &= \tfrac{1}{6}\left[\varrho_hr^2 + (1-\varrho_h)(3-2/r)\right], \label{eq:hydrostatic-solution-psi}
    \end{align}
\end{subequations}
where $\varrho_h$ is given by \eqref{eq:density-ratio} for the case $r_C=1$. When $r<1$, $\varrho=1$ so the singular term in \eqref{eq:hydrostatic-solution-psi} vanishes. Later, we use $\nu_h$, which is defined in a similar way to $\varrho_h$ from \eqref{eq:piecwise-viscosity} using the kinematic viscosity ratio $\nu$, again for the case $r_C=1$. The pressure at the centre of Earth is $P^*$. These equations are valid for all $r \le \curlyr$. Using $\boldsymbol{g}_h=-\Grad\Psi_h$, we can write the magnitude of gravity in the hydrostatic state, $g_h(r) = \tfrac{1}{3}\left[\varrho_h r + (1-\varrho_h)/r^2\right]$, so $g_h=1/3$ at the CMB. 

\subsection{Linearized equations}

To linearise the governing equations and boundary conditions, we write all independent variables in terms of a perturbation to the hydrostatic solution. Perturbation quantities are denoted by a $\Tilde{\,}$, except in the cases of the velocity $\vel$ and the tidal potential $\psi$ that are both zero in the hydrostatic reference state. The expansions are then substituted and terms quadratic in the perturbation are neglected.

The linearised, dimensionless boundary conditions are
\begin{subequations}
    \label{eq:boundary-conditions-linearised}
    \begin{align}
        \label{eq:noslip-boundary-condition}
        \vel\vert_{\curlyr} &= \boldsymbol{0},\\
        \label{eq:velocity-continuous-boundary-condition}
        \jump{\vel}_{1} &= \boldsymbol{0},\\
        \label{eq:stress-continuous-boundary-condition}
        \jump{-\pert{r}_C\fd{P_h}{r}\rhat + \pert{\stressten}\cdot\rhat}_1&=\boldsymbol{0}\\
        \label{eq:kinematic-boundary-condition}
        \partial_t{\pert{r}_C} - \vel|_{1}\cdot\rhat &= 0,
    \end{align}
\end{subequations}
where we have taken $\vel_S=\boldsymbol{0}$ and introduced the notation $\jump{q}_1 = q_O(r=1) - q_C(r=1)$ to represent the jump in any quantity, $q(\rpos)$, across the CMB.

In terms of perturbation quantities, the system \eqref{eq:governing-system-nondimensional} becomes
\begin{subequations}
    \label{eq:governing-system-perturbations}
    \begin{align}
        \label{eq:incompressibility-nondimensional}
        \Div\vel &= 0,\\
        \label{eq:momentum-equation-nondimensional}
        \pd{\vel}{t} &= \frac{\nu}{\re}\delsq\vel - \frac{\Grad \pert{P}}{\varrho_h} - \Grad\left(\pert{\Psi} - \tidepot\right),\\
        \label{eq:gravity-equation-nondimensional}
        \delsq\pert{\Psi} &= (1-\roc)\pert{r}_C\,\dirac(r-1),
    \end{align}
\end{subequations}
where $\dirac()$ is the non-dimensional Dirac delta function. Here we have collapsed the density perturbation associated with the interface displacement $\pert{r}_C$ onto $r=1$.  We are left with a linear, coupled problem for $\pert{\Psi},\,\pert{P},\,\vel.$

\subsection{Gravitational potential and pressure}
\label{sec:potential-solution}

By taking the divergence of \eqref{eq:momentum-equation-nondimensional} and using the continuity equation \eqref{eq:incompressibility-nondimensional}, we obtain
\begin{equation}
    \label{eq:pressure-equation-nondimensional}
    \delsq\left[\pert{P}+ \varrho_h(\pert{\Psi}-\tidepot)\right] = 0.
\end{equation}
This equation is valid within each subdomain, but not on the $r=1$ boundary between them.

Equations \eqref{eq:gravity-equation-nondimensional} and \eqref{eq:pressure-equation-nondimensional} form a linear system in $\pert{P},\pert{\Psi},\pert{r}_C$. Using the separable eigenfunctions as ansatz,
\begin{subequations}
    \label{eq:spherical-harm-ansatz}
    \begin{align}
        \label{eq:sphere-harm-ansatz-rC}
        \pert{r}_C(\theta,\phi,t) &=\frac{1}{2}\lmsumh \tidepot\elem r\elem \sph{\ell}{m}(\theta,\phi + \Omega t) + \cc,\\
        \label{eq:grav-ansatz-nd}
        \pert{\Psi}(\rpos,t) &=\frac{1}{2}\lmsumh \tidepot\elem \Psi\elem (r)\sph{\ell}{m}(\theta,\phi + \Omega t) + \cc,\\
        \pert{P}(\rpos,t) &=\frac{1}{2}\lmsumh \tidepot\elem P\elem (r)\sph{\ell}{m}(\theta,\phi + \Omega t) + \cc,
        \end{align}
\end{subequations}
where $r\elem$ is a set of constants and $P\elem,\Psi\elem$ are sets of radial eigenfunctions at each degree and order.  Linearity of the problem requires that the solution be linearly proportional to the magnitude of the forcing, and so we include a factor of $\tidepot\elem$ in each harmonic expansion.

With this ansatz, the solution of \eqref{eq:gravity-equation-nondimensional} for the radial functions is
\begin{equation}
    \label{eq:perturbed-potential-solution}
    \Psi\elem (r) = -\frac{(1-\roc)r\elem }{2\ell + 1}
    \begin{cases}
        r^\ell & r<1,\\
        r^{-\ell-1} & 1<r<\curlyr.
    \end{cases}
\end{equation}
Turning to equation~\eqref{eq:pressure-equation-nondimensional}, it is convenient to define $\pert{\Phi}\equiv\pert{P}/\varrho_h + \pert{\Psi} - \tidepot$. Then, requiring boundedness at $r=0$, the general solution is
\begin{equation}
    \label{eq:total-potential-component-definition}
    \Phi\elem  = (\ell+1)
    \begin{cases}
        A\elem r^\ell & r<1,\\
        B\elem r^\ell- \frac{\ell}{\ell+1} C\elem r^{-\ell-1} & 1<r<\curlyr,
    \end{cases}
\end{equation}
where $A\elem,B\elem,C\elem$ are sets of unknown constants. It is convenient for the next section to rewrite $\Grad\pert{\Phi}$ as a poloidal field,
\begin{equation}
    \label{eq:poloidal-grad-phi}
    \Grad\pert{\Phi} = \frac{1}{2}\lmsumh\tidepot\elem \Curl\Curl\left[\rhat\sph{\ell}{m}(\theta,\phi+\Omega t)\begin{cases}
        A\elem r^{\ell+1} & r<1,\\
        B\elem r^{\ell+1}+C\elem r^{-\ell} & 1<r<\curlyr
    \end{cases} \right] + \cc,
\end{equation}
where we have again introduced a factor of $\tidepot\elem$ into the ansatz.

\subsection{Velocity}
\label{sec:velocity-solution}
The momentum equation \eqref{eq:momentum-equation-nondimensional} can be rewritten by substituting the purely poloidal field $\Grad\pert{\Phi}$. This motivates a poloidal--toroidal decomposition of the velocity, 
\begin{multline}
    \label{eq:velocity-components}
    \vel(\rpos,t) = \frac{1}{2}\lmsumh\tidepot\elem\left\{\Curl\Curl\left[\rhat\,S\elem (r)\sph{\ell}{m}(\theta,\phi + \Omega t)\right]\right. + \\ \left.\Curl \left[\rhat\,T\elem (r)\sph{\ell}{m}(\theta,\phi + \Omega t)\right]\right\} + \cc.
\end{multline}
The forcing $\Grad\pert{\Phi}$ is poloidal and there are no mode-coupling terms in \eqref{eq:momentum-equation-nondimensional}, so the toroidal velocity components $T\elem(r)\equiv0$ for all $\ell, m$. Using this simplification, equation~\eqref{eq:momentum-equation-nondimensional} becomes
\begin{equation}
    \label{eq:poloidal-part-ode}
    \frac{\nu_h}{\re}r^2 \fd{^2S\elem }{r^2}-\left[im\Omega r^2+\frac{\nu_h}{\re}\ell(\ell+1)\right]S\elem (r) =
    \begin{cases}
        A\elem r^{\ell+3} & r<1,\\
        B\elem r^{\ell+3}+C\elem r^{-\ell+2} & 1<r<\curlyr.
    \end{cases}
\end{equation}
This is related to Bessel's equation. Defining $$q_C^m \equiv \sqrt{-im\Omega\re},\qquad q_O^m \equiv \sqrt{{-im\Omega\re}/{\noc}},$$ the solution of \eqref{eq:poloidal-part-ode} subject to boundedness at the origin is
\begin{equation}
    \label{eq:solution-S}
    S\elem(r) = 
    \begin{cases}
        -\frac{A\elem}{im\Omega}r^{\ell+1} + D\elem r^\frac{1}{2}J_{\ell+\frac{1}{2}}(q^m_Cr) & r<1,\\
        -\frac{B\elem}{im\Omega}r^{\ell+1} + E\elem r^\frac{1}{2}J_{\ell+\frac{1}{2}}(q^m_Or) -\frac{C\elem}{im\Omega}r^{-\ell} + F\elem r^\frac{1}{2}J_{-\ell-\frac{1}{2}}(q^m_Or) & 1<r<\curlyr.
    \end{cases}
\end{equation}
Here $A\elem,B\elem,C\elem,D\elem,E\elem,F\elem$ are sets of constants to be determined using the boundary conditions, and $J_n$ denotes the Bessel function of the first kind, of order $n$.

The result \eqref{eq:solution-S}, substituted into equation~\eqref{eq:velocity-components} with $T\elem=0$, gives the velocity components
\begin{subequations}
    \label{eq:velocity-components-S}
    \begin{align}
        \vel\cdot\rhat &= \frac{1}{2}\lmsumh\tidepot\elem\frac{\ell(\ell+1)}{r^2}S\elem\sph{\ell}{m} + \cc,\\
        \vel\cdot\that &= \frac{1}{2}\lmsumh\tidepot\elem\frac{1}{r}\fd{S\elem}{r}\pd{\sph{\ell}{m}}{\theta} + \cc,\\
        \label{eq:velocity-components-S-phi}
        \vel\cdot\phat &= \frac{1}{2}\lmsumh\tidepot\elem\frac{1}{r\sin\theta}\fd{S\elem}{r}\pd{\sph{\ell}{m}}{\phi} + \cc.
    \end{align}
\end{subequations}
Equations~\eqref{eq:velocity-components-S} with the radial structure \eqref{eq:solution-S} are used in the next section to write the boundary conditions \eqref{eq:boundary-conditions-linearised} in terms of the unknown constants.

\subsection{Application of boundary conditions}

Application of the boundary conditions is intricate in its details, which are provided in Appendix~\ref{app:BCs}. It is conceptually straightforward, however, and proceeds as follows. At each degree--order pair $\ell,m$, the solution is characterised by a set of seven unknown constants $A\elem,B\elem,C\elem,D\elem,E\elem,F\elem$ and $r\elem$. The boundary conditions provide seven constraints: each vector condition~(\ref{eq:boundary-conditions-linearised}a--c) can be written in terms of a radial part and a tangential part, giving six equations; the seventh comes from \eqref{eq:kinematic-boundary-condition}. These equations are expressed using the solutions of \S\ref{sec:potential-solution} and \S\ref{sec:velocity-solution}.

The resulting system of equations can be written in terms of the multiplication of a $7\times7$ matrix with a vector of unknowns at each $\ell,m$ pair (see eq.~\eqref{eq:boundary-condition-linear-system} in Appendix~\ref{app:BCs}).  Except in the inviscid case of $\nu_C,\nu_O\to0$, this system must be inverted numerically. At low Reynolds number, it is readily solved with standard linear-algebra software.  However, as the Reynolds number becomes large, the boundary-condition matrix becomes increasingly ill-conditioned due to the exponential behaviour of Bessel functions with complex argument, and inversion requires floating-point calculations with high precision.  At $\Omega\re\gg1$, the required floating point precision in terms of the number decimal digits $N_d$ is overestimated by a $\sim$10\% safety factor by 
\begin{equation}
    N_d \sim \texttt{round}\left(0.6\,\sqrt{\Omega\re}\right).
\end{equation}
For $\Omega=1$, $\re = \{10^5,\,10^7,\,10^9\}$ this gives $N_d = \{190,\,1898,\,18974\}$.  The base-10 logarithm of the condition number of the boundary-condition matrix gives a posterior estimate of the digits that were actually required; in these three cases it is $\{170,\,1683,\,16809\}$. We use the Advanpix Multiprecision Toolbox for \textsc{Matlab} that enables the user to specify the precision of calculations \citep{mct2015}.

\section{An analytical solution for inviscid flow}
\label{sec:inviscid-solution}

Before proceeding to investigate solutions obtained by inversion of the boundary-condition matrix, we consider the case of infinite Reynolds number, where both the core and the BMO are inviscid.  While this approximation is readily justified for the core on physical grounds, it is less clear whether the basal magma ocean was inertially dominated throughout its lifetime.  We discuss viscous effects in \S\ref{sec:viscous-solutions} below, and here develop the fully inviscid case as a means to gain insight into the broad structure and properties of the solution.

\begin{figure}
    \centering
    \includegraphics[width=\linewidth]{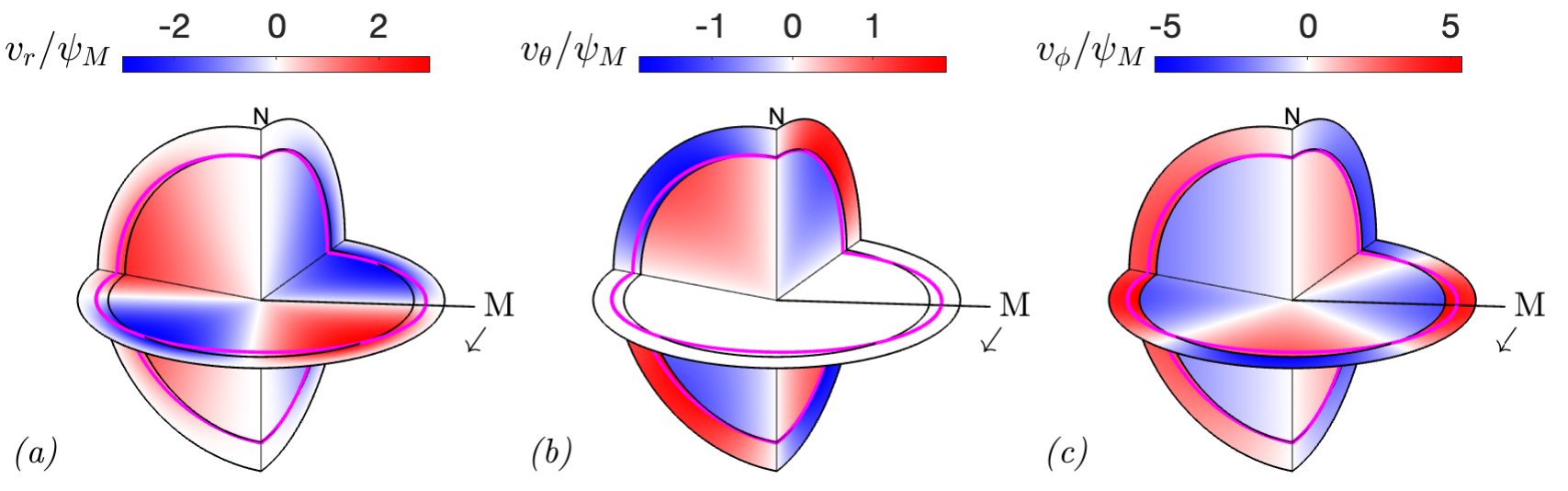}
    \caption{Velocity components, shown in the planetary reference frame, of the $2,2$ inviscid solution with  $\curlyr=1.2$,  $\roc=1/2$ and $\Omega=\Omega_*/2\approx0.088$ (marked \textbf{A} in Fig.~\ref{fig:inviscid-maps}). The $z$ axis is upward, toward the North pole.  The position of the Moon (not to scale) is marked as M. Black circle-segments mark the outer boundary of the core and BMO. Magenta lines show the CMB ellipsoid with radial exaggeration. \textit{(a)} Radial component $v_r$. \textit{(b)} Meridional component $v_\theta$. \textit{(c)} Zonal component $v_\phi$.}
    \label{fig:inviscid-solution}
\end{figure}

The inviscid case is a singular perturbation of the model; boundary conditions on the tangential components of the flow (and stress) cannot be enforced. All of the $D\elem$, $E\elem$, $F\elem$ coefficients are zero and the boundary-condition matrix is analytically invertible for $A\elem,B\elem,C\elem$ and $r\elem$. The result is
\begin{subequations}
\label{eq:inviscid-analytical-coefs}
\begin{align}
    \label{eq:inviscid-A}
    A\elem &=\frac{(1-\roc)/(\ell+1)}{\frac{\ell\left(1-\roc\right)}{m^2\Omega^2}\left(\frac{1}{3}-\frac{1-\roc}{2\ell+1}\right)  - \frac{\roc}{1-\curlyr^{-2\ell-1}}\left(\curlyr^{-2\ell-1}+\frac{\ell}{\ell+1}\right)- 1},\\
    \label{eq:inviscid-BCr}
    B\elem&=\frac{1}{1-\curlyr^{2\ell+1}}A\elem,\qquad 
    C\elem =\frac{1}{1-\curlyr^{-2\ell-1}}A\elem,\qquad
    r\elem=\frac{\ell(\ell+1)}{m^2\Omega^2}A\elem.
\end{align}
\end{subequations}
These coefficients can be used to construct $S\elem$ via equation~\eqref{eq:solution-S}, and hence the velocity components via equations~\eqref{eq:velocity-components-S}.  

Plots of the velocity components for the 2,2 inviscid tide are shown in Figure~\ref{fig:inviscid-solution}.  Flow within the core is identical to the canonical elliptical vortex solution (confirmed analytically in \S\ref{sec:compare-elliptical-vortex}). Flow within the BMO is dominantly tangential: zonal ($v_\phi$) and meridional ($v_\theta$). The radial component is limited by the thinness of the BMO. The zonal and meridional flows are driven by the tidal rise and fall of the CMB. Where the CMB is rising, the tangential flow of the BMO must diverge; where the CMB is falling, the tangential flow converges. This is the most fundamental characteristic of the tidal response in the core--BMO system.

\subsection{Ellipticity, the equilibrium tide, and tidal resonance} \label{sec:ellipticity}

The maximum height of the 2,2 lunar tide is quantified by the ellipticity $\ellipticity$. A larger value of $\ellipticity$ at a given frequency $\Omega$ is associated with faster flow.  This means stronger shear in the core, and hence greater potential for instability, and it means faster tangential flow in the BMO (at a given BMO thickness $\curlyr-1$).  Ellipticity is therefore the most important metric of the solution.

\begin{figure}
    \centering
    \includegraphics[width=\linewidth]{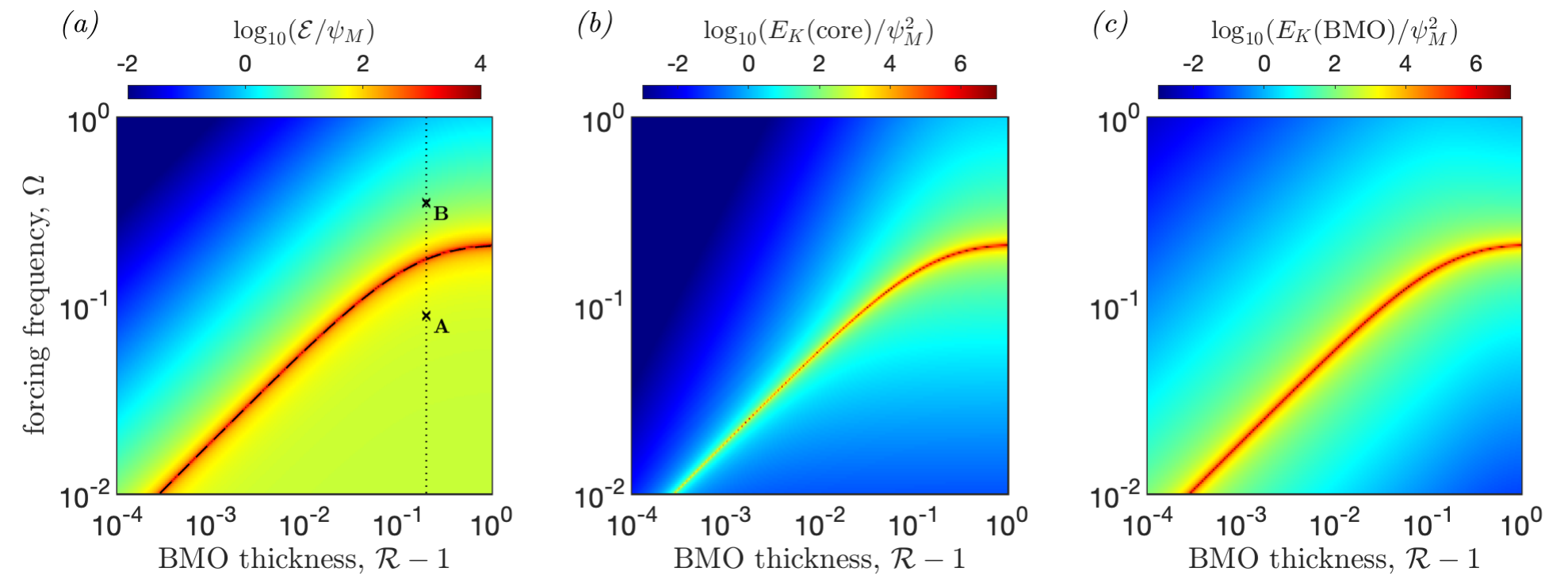}
    \caption{Aspects of the inviscid solution for the $2,2$ lunar tide with $\roc=1/2$. Colour scales are truncated at the upper end because the resonance is a singularity. \textit{(a)} The normalised ellipticity $\ellipticity/\tidepotm$ calculated with eq.~\eqref{eq:inviscid-ellipticity}.  The resonance trend of eq.~\eqref{eq:inviscid-resonance-trend} is plotted as a dashed line. The points marked with $\boldsymbol{\times}$  denote frequencies $\Omega_*/2$ (point~\textbf{A}) and $2\Omega_*$ (point~\textbf{B}). The flow at point \textbf{A} is shown in Fig.~\ref{fig:inviscid-solution}. \textit{(b)} The normalised and dimensionless kinetic energy within the core, corresponding to eq.~\eqref{eq:KE-inviscid-core}. \textit{(c)} The normalised and dimensionless kinetic energy within the BMO. Both kinetic-energy panels are calculated with eq.~\eqref{eq:kinetic-energy-radial-integral}.}
    \label{fig:inviscid-maps}
\end{figure}

\begin{figure}
    \centering
    \includegraphics[width=\linewidth]{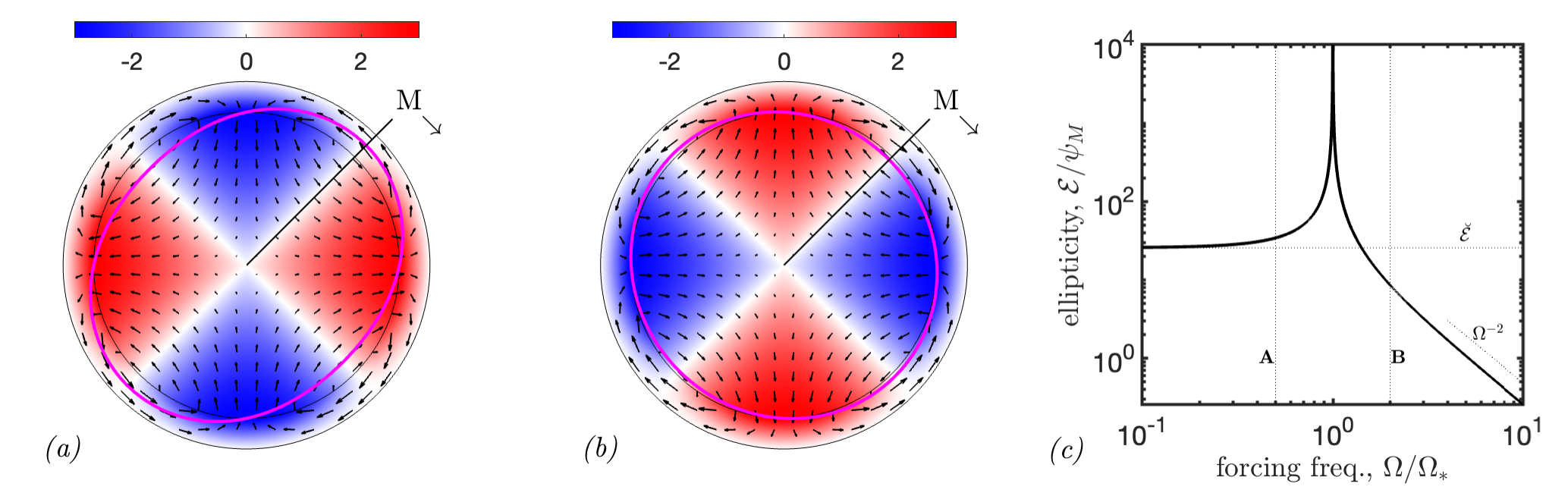}
    \caption{Equatorial flow portraits, in the planetary reference frame, of the $2,2$ inviscid solution with  $\curlyr=1.2$ and  $\roc=1/2$. Colour shows $v_r$ and vectors show the full velocity field. Thin black lines mark the CMB and the outer radius of the BMO. Magenta ellipse depicts $r_C$ with radial exaggeration (the same exaggeration factor is used in both portraits). \textit{(a)} The case marked \textbf{A} in Fig.~\ref{fig:inviscid-maps}a where $\Omega\approx0.09$. \textit{(b)} The case marked \textbf{B} in Fig.~\ref{fig:inviscid-maps}a where $\Omega \approx 0.35$. \textit{(c)} Normalised ellipticity plotted as a function of $\Omega/\Omega_*$.  Frequencies of \textbf{A} and \textbf{B} are vertical dotted lines and the normalised equilibrium ellipticity $\equil{\ellipticity}/\tidepotm$ is plotted as a horizontal dotted line.}
    \label{fig:inviscid-equatorial}
\end{figure}

With the result \eqref{eq:inviscid-analytical-coefs}, we can reconstruct the interface perturbation $\pert{r}_C(\theta,\phi,t)$ of the $2,2$ lunar tide. Using this, the dimensionless semi-major ($a$) and semi-minor ($b$) axes of the equatorial-plane CMB ellipse are $1\pm|\pert{r}_C|$. In terms of $a$ and $b$, the ellipticity is defined as $\ellipticity=(a^2-b^2)/(a^2+b^2)$.  Because $|\pert{r}_C|$ is much smaller than unity, we can make a linear approximation of the ellipticity to 
\begin{equation}
    \ellipticity\sim a-b = 2|\pert{r}_C|, 
    \qquad(|\pert{r}_C|\ll1;\;\ell=m=2\text{ mode only}). 
    \label{eq:ellipticity-rc}
\end{equation}
Hence, using \eqref{eq:inviscid-BCr} and \eqref{eq:sphere-harm-ansatz-rC} evaluated at the equator ($\theta=\pi/2$), we find that 
\begin{equation}
    \label{eq:inviscid-ellipticity}
    \ellipticity\sim 9|A_2^2|\tidepotm/\Omega^2\qquad\qquad\text{(inviscid)}.
\end{equation}
This ellipticity of the CMB is plotted in Figure~\ref{fig:inviscid-maps}a as a function of forcing frequency $\Omega$ and BMO thickness $\curlyr-1$ (note that most figures show the ellipticity normalised as $\ellipticity/\tidepotm$; this ratio is large but $\tidepotm \ll 1$). We see that the ellipticity response of the CMB has three features: low- and high-frequency regimes separated by a resonance trend.  We discuss these two regimes in terms of the solutions at points \textbf{A} and \textbf{B} in Figure~\ref{fig:inviscid-maps}a, corresponding to the equatorial-plane portraits in Figure~\ref{fig:inviscid-equatorial}a and b, respectively. The profile of $\ellipticity$ along the dotted line in Fig.~\ref{fig:inviscid-maps}a is shown in Fig.~\ref{fig:inviscid-equatorial}c.

The low-frequency regime forms a plateau in ellipticity $\ellipticity$ at small $\Omega$. This is the equilibrium tide, in which the pressure adjusts so that the tidal forcing is balanced by hydrostatic and self-gravitational restoring effects. Figure~\ref{fig:inviscid-equatorial}a shows an equatorial-plane portrait of the flow regime. The magenta ellipse (radially exaggerated) has a lag $\Delta\phi$ of zero with respect to the position of the Moon.

Figure~\ref{fig:inviscid-equatorial}c shows that as $\Omega\to0$, the ellipticity \eqref{eq:inviscid-ellipticity} is asymptotic to
\begin{equation}
    \label{eq:inviscid-equil-ellipticity}
    \equil{\ellipticity} \equiv \lim_{\Omega\to0}\ellipticity = \frac{6(2\ell+1)}{\tfrac{2}{3}(\ell-1)+\roc}\tidepotm.
\end{equation}
The lunar tidal potential is predominantly $\ell=2$, which gives $\equil{\ellipticity} = 30\tidepotm/(2/3+\roc).$ When $\roc=1/2$, as used throughout this paper, $\equil{\ellipticity}\approx25.7\tidepotm$; this is the value attained in the lower-right corner of Fig.~\ref{fig:inviscid-maps}a. Because the equilibrium tide represents a quasi-static balance, it holds in the limit of $\Omega\to 0$ even when viscosity is non-zero (as in Fig.~\ref{fig:viscous-Re-scaling}c, below).

The high-frequency regime occurs at frequencies larger than the resonance trend. This regime is exemplified by the conditions at point~\textbf{B} in Fig.~\ref{fig:inviscid-maps}a, which give a flow shown in the equatorial portrait of Fig.~\ref{fig:inviscid-equatorial}b. The tidal ellipse in the high-frequency regime is not aligned with the Moon, but rather lags it at $\Delta\phi = \pi/2$ (magenta curve; Fig.~\ref{fig:inviscid-equatorial}b). This represents an interfacial gravity tidal wave. Here the tidal body force balances inertia to create a forced wave.  This balance results in ellipticity scaling with $\Omega^{-2}$ for $\Omega\gg\Omega_*$, as shown in Figure~\ref{fig:inviscid-equatorial}c.

Figure~\ref{fig:inviscid-maps}a shows that the low- and high-frequency regimes join at a resonance, where the ellipticity is singular (the colour scale is truncated; see also Fig.~\ref{fig:inviscid-equatorial}c).  The forcing frequency $\Omega_*$ associated with this resonance corresponds to the denominator of \eqref{eq:inviscid-A} going to zero,
\begin{equation}
    \label{eq:inviscid-resonance-trend}
    \Omega^2_* = \frac{\Omega_\text{Kel}^2}{m^2}
    \left[1+\frac{3\roc}{2(\ell-1)}\right]\left[1 + \frac{\roc(2\ell+1)}{(1-\roc)(\ell+1)\left(1-\curlyr^{-2\ell-1}\right)}\right]^{-1} \quad\text{(inviscid)}.
\end{equation}
We have written the resonance frequency in terms of the eigenfrequencies $\Omega_\text{Kel}^2 = 2\ell(\ell-1)/3(2\ell+1)$ of an inviscid, self-gravitating sphere (first obtained by Kelvin \citep{lamb1924hydrodynamics}). The limit of a zero-density BMO ($\roc=0$) recovers Kelvin's result. The $2,2$ case is plotted as a dashed line in Fig.~\ref{fig:inviscid-maps}a. 

In physical terms, this resonance occurs when the tidal forcing frequency coincides with the natural oscillation frequency of the $2,2$ mode on the CMB. In other words, it occurs when the tidal potential advances at exactly the wave-speed of an equatorial, interfacial gravity wave \citep{tyler2014ResonantTidal}.  In this case, the tidal force enhances the amplitude of the interfacial gravity wave. In the absence of viscous dissipation, this resonance is manifest as a singularity. Laminar viscous effects (\S\ref{sec:viscous-solutions}) damp the singularity, and the combination of turbulence, viscous dissipation and magnetic effects would promote a non-linear saturation \citep{barker2014non}.  Moreover, the frequency of the unforced wave (the eigenmode) is modified by the Coriolis force, which is not accounted for here.  We return to this topic in the discussion section (\S\ref{sec:discussion}).

\subsection{Potential and kinetic energy}

The potential energy of a solution arises from the displacement of the interface between the core and the BMO.  At any point within the domain, the dimensional potential energy per unit volume (in excess of the static potential energy) is given by $e_P(\rpos) = \Delta\rho(\rpos)g(\rpos)(r-R_C)$, where $\Delta\rho(\rpos)$ is the density difference from the static case. The region of non-zero $\Delta\rho$ is the volume $V_P$ between the tidally displaced CMB and its reference position. Since the CMB displacement is formally small, we can linearise the integral of the potential energy within this volume; an analytical expression~\eqref{eq:potential-energy} is given in Appendix~\ref{app:energy}. 

Using \eqref{eq:potential-energy} to compute the dimensionless potential energy of the inviscid solution for the $2,2$ mode results in
\begin{equation}
    \label{eq:EP-inviscid}
    E_P \sim (1-\roc)N_2^2\ellipticity^2/144\qquad\text{(inviscid)},
\end{equation}
where $N\elem$ is a mode-specific normalisation factor defined in eq.~\eqref{eq:normalisation-factor}. Considering equation~\eqref{eq:EP-inviscid} in the case of $\Omega\to0$, using $\equil{\ellipticity}$ from eq.~\eqref{eq:inviscid-equil-ellipticity}, and taking $\roc=1/2$, we obtain the potential energy of the equilibrium tide, $\equil{E}_P/\tidepotm^2\approx138.3$.  The dimensional potential energy is obtained by multiplying by $[E]$. 

A plot of potential energy~\eqref{eq:EP-inviscid} as a function of frequency and BMO thickness (not shown) closely mirrors the plot of ellipticity in Fig.~\ref{fig:inviscid-maps}a: it has the plateaux at small $\Omega$, the resonance trend, and a (steeper) decline toward large $\Omega$. 

The kinetic energy arises from the motion of the fluid throughout the domain.  At any point within the domain, the dimensional kinetic energy per unit volume is given by $e_K(\rpos) = \rho\vel\cdot\vel/2$. The total kinetic energy within a radius $r_i \le r \le r_o$ is then obtained by integrating $e_K$ over this volume. In appendix~\ref{app:energy} we present an analytical expression~\eqref{eq:KE-solid-angle} for $\solidangle{e}_K(r)$, which is the kinetic energy density integrated over $4\pi$ steradians at radius $r$.  With this we can compute the total kinetic energy of the core $E_K(\text{core}) = \int_0^1 r^2\solidangle{e}_K\,\infd r$ and the BMO $E_K(\text{BMO}) = \int_1^\curlyr r^2\solidangle{e}_K\,\infd r$, shown in Fig.~\ref{fig:inviscid-maps}b and c, respectively. The kinetic energy of the core in the inviscid solution is obtained by combining \eqref{eq:inviscid-analytical-coefs} with \eqref{eq:solution-S} to calculate $S\elem$, and using this in \eqref{eq:KE-solid-angle}. For the $2,2$ inviscid mode, the total dimensionless kinetic energy of the core is
\begin{equation}
    \label{eq:KE-inviscid-core}
    E_K(\text{core}) = 5N_2^2\ellipticity^2\Omega^2/216 \qquad (\text{inviscid}).
\end{equation}
For $\Omega$ sufficiently far below resonance, we can approximate $\ellipticity \sim \equil{\ellipticity} \approx 25.7\tidepotm$.  Therefore, the kinetic energy scales with $\Omega^2$ as $\Omega\to0$. Above resonance, $\ellipticity$ scales as $\Omega^{-2}$ for $\Omega\to\infty$, and hence so too does $E_K$. The dimensional kinetic energy is obtained by multiplying by $[E]$.

\subsection{Comparison with the canonical elliptical vortex}
\label{sec:compare-elliptical-vortex}

The canonical elliptical vortex has been studied extensively in the context of tidally driven astrophysical and geophysical flows \citep[e.g.,][]{pierrehumbert1986universal, kerswell2002elliptical, barker2016non-linear-919}. It arises as a strain-flow perturbation that rotates relative to the underlying solid-body flow, and is an exact solution of the incompressible, inviscid Euler equations.  This situation appears to arise in the context of the $2,2$ tide at radii $r<1$ within the core of our model.  Using the inviscid solution \eqref{eq:inviscid-analytical-coefs} to compute the velocity components \eqref{eq:velocity-components-S} and projecting these into a Cartesian basis, we obtain 
\begin{equation}
    \label{eq:core-inviscid-cartesian}
    \begin{bmatrix}
        v_x \\ v_y \\ v_z
    \end{bmatrix} = -\ellipticity\Omega
    \begin{bmatrix}
        \sin 2\Omega t & \cos 2\Omega t & 0 \\
        \cos 2\Omega t & -\sin 2\Omega t & 0 \\
        0 & 0 & 0
    \end{bmatrix}
    \begin{bmatrix}
        x \\ y \\ z
    \end{bmatrix} \qquad(\text{inviscid; for }|\pos|<1),
\end{equation}
where $\ellipticity$ is given by equation~\eqref{eq:inviscid-ellipticity}. This solution, shown in Figure~\ref{fig:inviscid-solution}, is linear in the Cartesian coordinates and has zero vorticity and uniform strain rate.  The solution for $1<|\pos|<\curlyr$ has two parts, with coefficients $B\elem$ and $C\elem$ respectively.  The first of these is proportional to \eqref{eq:core-inviscid-cartesian}, whereas the second part differs, and has non-zero $v_z$. For $\curlyr-1\ll1$, the magnitude of these two parts is identical.

The result \eqref{eq:core-inviscid-cartesian} is identical to the canonical elliptical vortex solution in the planetary reference frame (a didactic derivation is provided by \cite{lereun2020}). It establishes a direct mathematical connection between the present two-layer tidal problem and the canonical elliptical-vortex solutions that underpin existing theories of tidally driven instability in planetary cores. In particular, it indicates that our inviscid solution is susceptible to the elliptical instability \citep[e.g.,][]{kerswell2002elliptical}, as discussed in \S\ref{sec:Ei-instabilty}.

\section{Viscous solutions}
\label{sec:viscous-solutions}

In this section we restore viscosity to the problem and examine its effects.  We first consider the full solution, and examine how it depends on viscosity via the Reynolds number $\re$.  We then briefly consider solutions for an inviscid core and viscous BMO---a limiting case in which $\noc \equiv  \nu_O/\nu_C \rightarrow \infty$.  This reduces the number of parameters by one. It is a useful context in which to examine the consequences of high viscosity in the BMO, such as might arise if it becomes a crystal-rich suspension. Finally, we calculate the rate of viscous dissipation, both in the core and the BMO from the general viscous model, and in the BMO for the simpler inviscid core and viscous BMO model.

\subsection{Viscous core \& BMO}

\begin{figure}
    \centering
    \includegraphics[width=\linewidth]{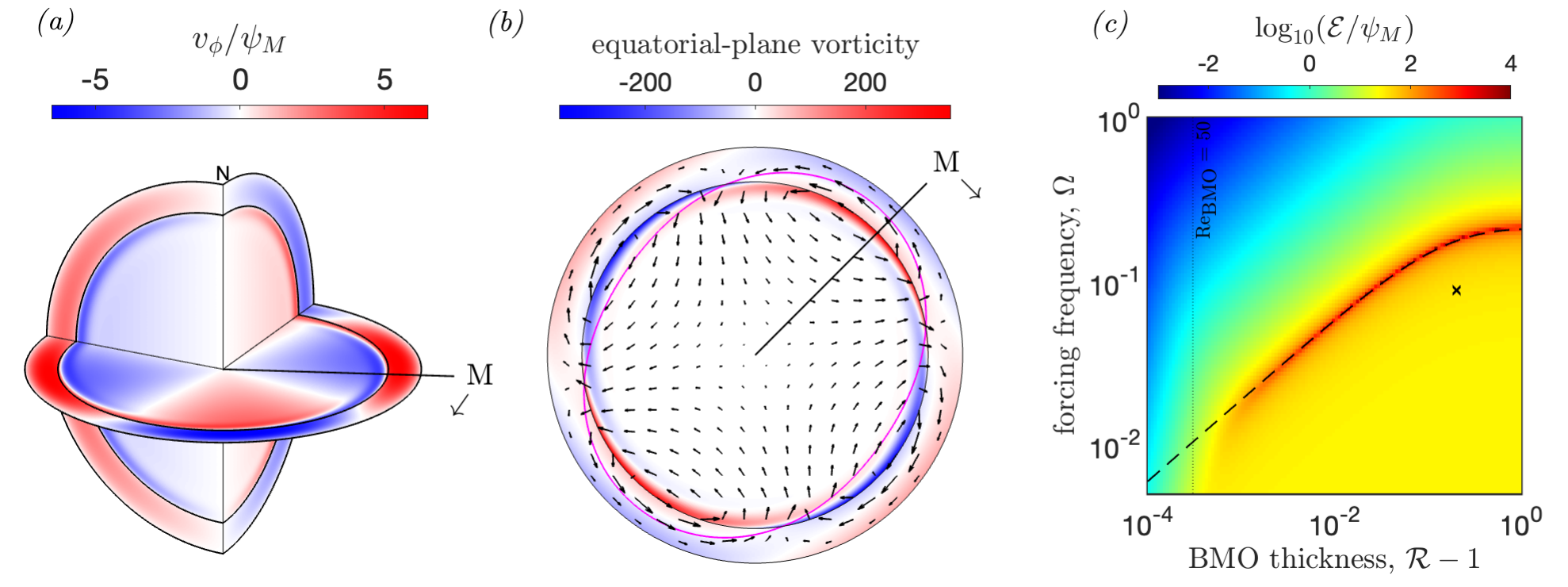}
    \caption{The viscous solution with $\roc=1/2$ and $\noc=20$. \textit{(a)} The zonal velocity component $v_\phi$ at $\re=10^4$, $\curlyr=1.2$ and $\Omega=\Omega_*/2$. \textbf{(b)} Vorticity $\zhat\cdot\Curl\vel$ in the equatorial plane ($\theta=\pi/2$). Vectors show the full flow field. Parameters as in panel~a.  The magenta line shows a radially exaggerated $r_C$, which lags the Moon by $\Delta\phi\approx-9^\circ$ due to viscosity. \textit{(c)} Map of normalised ellipticity in parameter space of forcing frequency and BMO thickness, for $\re=10^{10}$. The $\times$ marks the $\Omega,\curlyr$ conditions used in panels~a and b.}
    \label{fig:viscous-solution}
\end{figure}

Figure~\ref{fig:viscous-solution}a shows the zonal velocity component $v_\phi$ from a viscous solution with $\re=10^4$ and $\noc=20$. The corresponding plot of the inviscid flow is Fig.~\ref{fig:inviscid-solution}c.  Viscous boundary layers around $r=1$ and beneath $r=\curlyr$ are present and enable the interior flow to satisfy the no-slip boundary conditions.  Away from these boundary layers, in the interior of the core and the BMO, the flow is asymptotic to that of the inviscid case as $\re\to\infty.$

Vorticity is generated within the viscous boundary layers.  The vorticity in the equatorial plane is shown in Figure~\ref{fig:viscous-solution}b.  Bands of alternating sign are aligned with the tidal ellipse, shown as a magenta line (not to scale). In \S\ref{sec:BL-instability} we consider how this boundary layer might become unstable to turbulence.

Figure~\ref{fig:viscous-solution}c shows the ellipticity per forcing magnitude $\ellipticity/\tidepotm$ as a function of BMO thickness and forcing frequency. This panel is computed with $\re=10^{10}$ (the largest value at which it is practical to compute this solution array). The resonance trend is evident, though its amplitude is limited by the finite viscosity.  For large BMO thickness, the resonance trend follows the inviscid prediction (dashed line).  In this range, the viscous boundary layers at the top and bottom of the BMO do not interact.  But for $\curlyr-1 \lesssim 10^{-3}$, the resonance trend departs from the inviscid prediction and dissipates.  We can form a Reynolds number for the BMO as 
\begin{equation}
    \label{eq:BMO-Reynolds}
    \re_\text{BMO} \equiv (\curlyr-1)^2\,\re/{\noc}
\end{equation}
and with $\noc=20$, the departure from the inviscid resonance occurs at $\re_\text{BMO}\approx 50$. In this case, the BMO is thin enough that the boundary layers interact across its height.

\begin{figure}
    \centering
    \includegraphics[width=\linewidth]{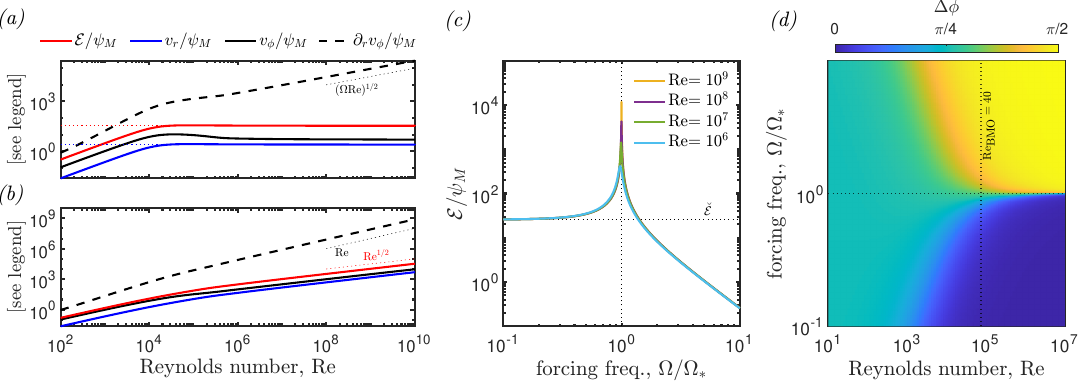}
    \caption{Scaling of the viscous solution with Reynolds number for $\curlyr=1.1$, $\roc=1/2$, $\noc=20$. \textit{(a)} Scaling of solution components with Re at $\Omega=\Omega_*/2$. Horizontal dotted lines represent the inviscid analytical solution. Velocity components are evaluated as the maximum value at the equator of the CMB ($r=1,\,\theta=\pi/2$). \textit{(b)} Scaling of solution components with Re at $\Omega = \Omega_*$, i.e., at resonance. Lines as in panel~a. \textit{(c)} Normalised ellipticity as a function of forcing frequency for different values of the Reynolds number. \textit{(d)} The phase lag $\Delta\phi$ between the Moon and the semi-major axis of the tidal ellipse in the equatorial plane. The inviscid solution has $\Delta\phi = 0$ for $\Omega/\Omega_*<1$ and $\Delta\phi = \pi/2$ otherwise.}
    \label{fig:viscous-Re-scaling}
\end{figure}

The Reynolds number evidently exerts an important control on boundary-layer thickness and properties of the resonance. Figure~\ref{fig:viscous-Re-scaling} illustrates the scaling of the solution with $\re$.  Panel~(a) plots solution metrics at non-resonant conditions.  For $\re\gtrsim10^6$, ellipticity and velocity components are in a regime that is independent of further increase in $\re$.  For $\ellipticity$ and $v_r$, the high-$\re$ regime is asymptotic to the inviscid solution shown as dotted lines; for $v_\phi$, the viscous boundary layer at $r=1$ creates a qualitative difference from the inviscid solution.  The thickness of the viscous boundary layer scales with with $(\Omega \re)^{-1/2}$ and hence the radial shear $\partial_rv_\phi$ in this layer scales as the inverse of this (dashed line; details in \S\ref{sec:BL-instability}).  The scalings shown in this panel hold everywhere except on the resonance trend.

Scaling of the solution with $\re$ is different at resonance.  This is shown in Fig.~\ref{fig:viscous-Re-scaling}b.  A high-$\re$ regime exists at resonance, but the velocity in this regime increases with $\re^{1/2}$. Because of this, the radial shear increases with $\re$. Panel~(c) shows that this increase is entirely localised to the resonance frequency $\Omega_*$; there is no increase away from $\Omega_*$. 

Figure~\ref{fig:viscous-Re-scaling}d shows that viscosity also affects the lag, $\Delta\phi$ (Fig.~\ref{fig:domain}), between the tidal bulge and the tidal potential.  In the inviscid case, the lag is zero for $\Omega\le\Omega_*$, or $\pi/2$ otherwise (Fig.~\ref{fig:inviscid-equatorial}).  But viscous effects cause the tidal bulge to lag by $\sim\pi/4$ at smaller $\re$.  The transition seems to be controlled by $\re_\text{BMO}$ (eq.~\eqref{eq:BMO-Reynolds}) because the BMO is more viscous and thinner than the core; hence, viscous effects in the BMO onset at higher $\re$. 

The importance of viscous effects in the BMO motivates us to consider a third variant of the model in which the core is inviscid.

\subsection{Inviscid core, viscous BMO}
\label{sec:ICVO-solution}

The BMO is a silicate melt that is more polymerised and hence more viscous than a liquid-iron alloy at the same pressure--temperature conditions.  Furthermore, progressive crystallisation of the BMO will reduce its thickness and may increase its viscosity through suspension of crystals. We therefore expect the Reynolds number of the BMO to decrease over time. At some sufficiently small Reynolds number, we expect viscous effects in the BMO to dampen the tidal response. 

\begin{figure}
    \centering
    \includegraphics[width=\linewidth]{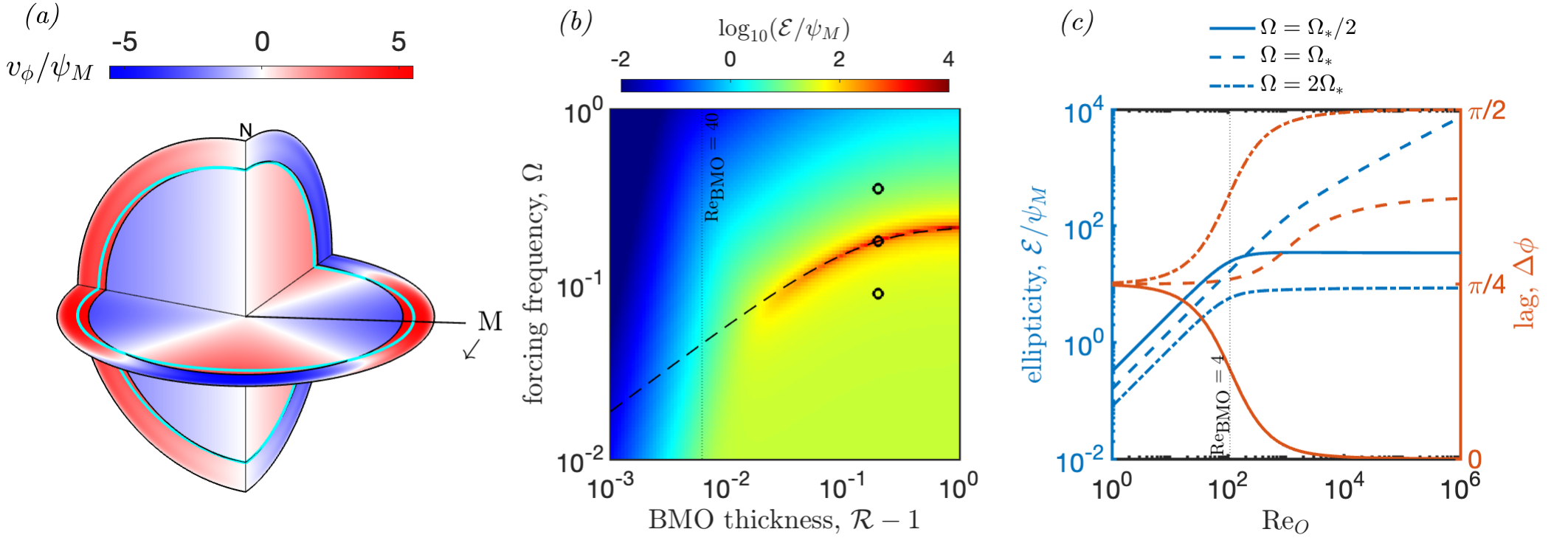}
    \caption{Solutions of the ICVO case. \textit{(a)} Zonal flow field $v_\phi$ with $\reo=10^4$, $\roc=1/2$ and $\curlyr=1.2$. Cyan line shows $r_C$ but is plotted with exaggerated ellipticity. The BMO flow satisfies no stress at $r=1$ and no slip at $r=\curlyr$. \textit{(b)} Normalised ellipticity as a function of BMO thickness and forcing frequency for $\reo=10^6$ with $\roc=1/2$. Dashed line marks the inviscid resonance from eq.~\eqref{eq:inviscid-resonance-trend}. Circles mark the conditions for the curves in panel~c. \textit{(c)} Ellipticity and lag as a function of $\reo$. Line style indicates the forcing frequency, as in the legend. }
    \label{fig:ICVO-plots}
\end{figure}

To investigate this, we reformulate the problem for an inviscid core and viscous BMO, which we refer to as the ICVO case.  As in the fully inviscid case of \S\ref{sec:inviscid-solution}, this is a singular perturbation under which we cannot enforce the condition of continuity of tangential velocity across $r=1$. Moreover, the continuity of shear stress at $r=1$ becomes a no-stress condition at the bottom of the BMO. The boundary-condition matrix is modified to accommodate this, and $D\elem=0$ such that we lose all dependence on $q_C^m$. Because $\nu_C=0$, $\noc$ and $\re$ become meaningless. Thus we make the replacement $\reo = \re/\noc$, which is independent of $\nu_C$, where $\reo$ is a scaled version of the core Reynolds number based on the length and velocity scales for the core flow (table~\ref{tab:parameters}) but using the viscosity of the BMO. Then the dynamically relevant Reynolds number for the BMO (eq.~\ref{eq:BMO-Reynolds}) can be expressed $\re_\text{BMO} \equiv (\curlyr-1)^2\,\reo$.

Figure~\ref{fig:ICVO-plots} shows solutions of the ICVO case.  The zonal flow field and CMB ellipsoid are plotted in panel~(a) for $\reo=10^4$.  Inviscid flow in the core has no viscous boundary layer at $r=1$; the no-stress condition in the BMO at $r=1$ also inhibits a boundary layer.  The no-slip condition at $r=\curlyr$ requires a boundary layer, and this is evident where it is not too thin to be seen.  Panel~(b) of this figure plots the ellipticity (per forcing magnitude) as a function of BMO thickness and frequency for $\reo=10^6$.  The pattern is much like the fully viscous case shown in Fig.~\ref{fig:viscous-solution}c: the resonance trend matches the inviscid prediction when the BMO thickness is sufficiently large.  The scaling with $\reo$ is shown in panel~(c).  Ellipticity is shown in blue and lag is shown in red; three line styles indicate frequencies below, at, and above resonance.  For all curves, a high-$\reo$ regime is apparent above $\reo \approx 10^3$. The vertical line marks $\re_\text{BMO}=4$, which corresponds to $\reo=100$ for $\curlyr=1.2$. The scaling of ellipticity and lag shown in this figure are qualitatively consistent with that of Fig.~\ref{fig:viscous-Re-scaling}, so for explanation we refer the reader back to the discussion of that figure.

\subsection{The rate of viscous dissipation}

Having incorporated viscosity into the model, we can now use our solutions to predict a rate at which the energy of the forcing is dissipated.  The rate of working against viscosity per unit volume is given by $\dot{e}_\nu = 2\rho\nu\strainrten:\strainrten$, where $\strainrten$ is the strain-rate tensor, defined in the usual way for incompressible flow. In Appendix~\ref{app:energy} we non-dimensionalise this and integrate it over the volume between two radii. To compute the total dissipation in the core, we integrate for $r\in[0,1]$ and for that of the BMO, $r\in[1,\curlyr].$ The integration is performed numerically by the trapezoidal rule, with a logarithmic spacing of nodes in the core providing high resolution of the integrand in the viscous boundary layer beneath the CMB. The BMO is sampled uniformly in radius.  Both integrals use 1000 nodes, which provides a result converged to 1\% or less in the BMO (the error in the core is about 100$\times$ smaller).

\begin{figure}
    \centering
    \includegraphics[width=0.95\linewidth]{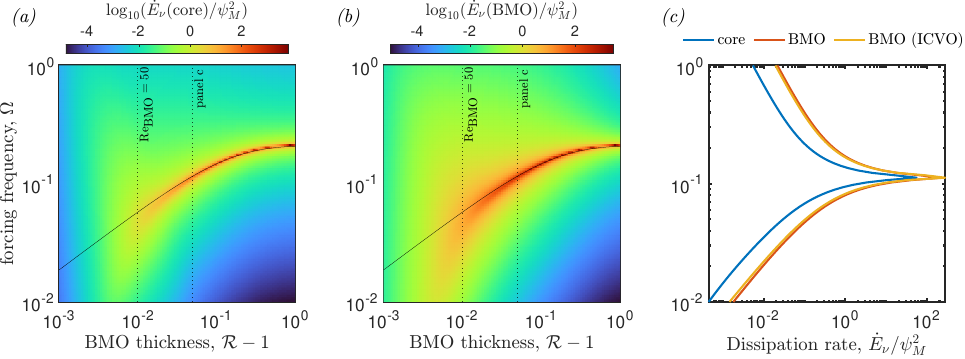}
    \caption{Rate of volume-integrated viscous energy dissipation $\dot{E}_\nu$ for $\re=10^7$ with $\roc=1/2,\,\noc=20$, per square forcing magnitude. \textit{(a)} Dissipation rate in the core. \textit{(b)} Dissipation rate in the BMO. \textit{(c)} Dissipation rate at $\curlyr=1.05$ for the fully viscous model and an ICVO model with $\reo = 10^{7}/20$.}
    \label{fig:dissipation-rate}
\end{figure}

The rate of dissipation is shown in Figure~\ref{fig:dissipation-rate}.  Panels~(a) and (b) represent a fully viscous case and plot the rate of dissipation in the core and BMO, respectively.  Panel~(c) compares these to the BMO dissipation in the ICVO case at $\curlyr=1.05$.  For the fully viscous case, we use Reynolds number $\re=10^7$ and the BMO viscosity is measured by $\noc = 20$. For the ICVO case, we use Reynolds number $\reo = \re/\noc = 5\times10^5$ (we consider dimensional parameter values and outcomes in the Discussion, below). The resonance trend is a locus of higher dissipation, as expected, but the off-resonance dissipation can remain high, particularly as the BMO thins.  Panel~c and, in particular, the comparison of its red and yellow curves, shows that dissipation in the BMO is almost insensitive to the (small) viscosity of the core.

The rate of dissipation, under any conditions except resonance, scales as the relevant Reynolds number to the $-1/2$ power.  At resonance, the dissipation is approximately independent of $\re$.

\section{Potential instabilities of the tidal flow}
\label{sec:instabilities}

Dynamo action requires flow that is sufficiently energetic and topologically complex. The former is assessed by the magnetic Reynolds number $\rem$ (\S\ref{sec:mag-Re}) while the latter is assessed by the propensity of the flow to generate turbulence via secondary instabilities. For tidally driven flows, there are two secondary instabilities that are most relevant. The first is an instability within the interior of the core called the elliptical instability (\S\ref{sec:Ei-instabilty}). The second is an instability of the boundary-layer shear flow between the core and the BMO (\S\ref{sec:BL-instability}).

\subsection{Magnetic Reynolds number} \label{sec:mag-Re}

\begin{figure}
    \centering
    \includegraphics[width=0.95\linewidth]{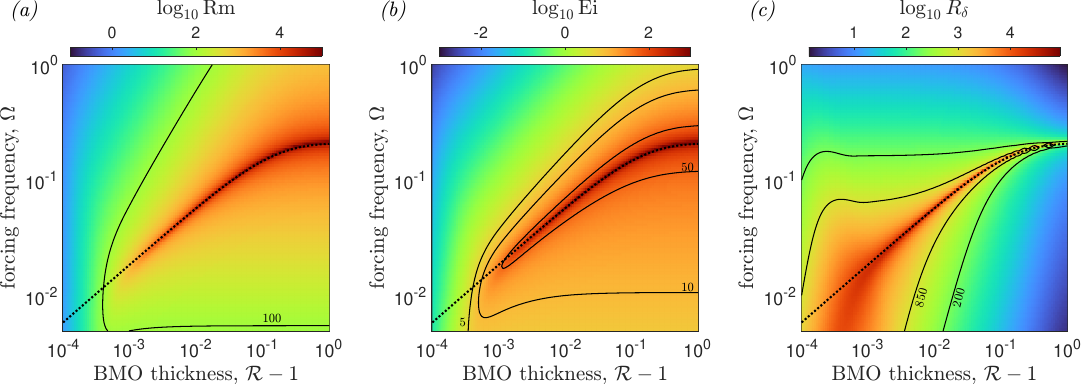}
    \caption{Dimensionless solution measures as a function of BMO thickness and forcing frequency. The fully viscous flow solution was computed at $\re=10^{10}$, with $\roc=1/2$, $\noc=20$ and $\tidepotm=2\times10^{-8}$. \textit{(a)} The magnetic Reynolds number of eq.~\eqref{eq:magnetic-reynolds}, evaluated from this flow solution using core properties corresponding to $\re\approx3.5\times10^{16}$ and $\pr=10^{-6}$. A contour indicating a critical value for $\rem$ is overlaid \citep{landeau2022sustaining}. \textit{(b)} The elliptical instability number of eq.~\eqref{eq:ell-instability-number}, evaluated from the $\re=10^{10}$ flow solution using core properties corresponding to $\re\approx3.5\times10^{16}$. Contours indicating possible values of $\Ei_\text{crit}$ are overlaid. \textit{(c)} The boundary-layer Reynolds number of eq.~\eqref{eq:R_delta_def}, calculated from the $\re=10^{10}$ flow solution and rescaled to $\re=10^{17}$. Contours indicating possible critical values for instability are overlaid.}
    \label{fig:rm-and-ei-plots}
\end{figure}

The magnetic Reynolds number $\rem$ characterises the relative rates of magnetic induction and magnetic diffusion within the flow. Induction scales with the characteristic speed of the flow, whereas diffusion scales with the magnetic diffusivity $\eta$.  Within a core driven by tides into a rotating ellipsoid with equatorial-plane ellipticity $\ellipticity$, the characteristic dimensional flow speed is \citep{landeau2022sustaining} 
\begin{equation}
    \label{eq:elliptical-flow-speed}
    \Uel \sim R_C\omega \ellipticity.
\end{equation}
We use this speed, together with the relevant length scale $R_C$, to form the magnetic Reynolds number, $\rem\equiv \Uel R_C/\eta = R_C^2\omega\ellipticity/\eta$. This can be re-written in dimensionless terms as 
\begin{equation}
    \label{eq:magnetic-reynolds}
    \rem = \Omega\,\re\,\pr\,\ellipticity,
    \qquad\text{where}\quad \pr \equiv \nu_C/\eta
\end{equation}
is the magnetic Prandtl number.  The ellipticity $\ellipticity$ is obtained from the solution for the primary flow (\S\ref{sec:ellipticity}) and has a non-trivial dependence on $\Omega$, $\re$ and $\curlyr$.  Figure~\ref{fig:rm-and-ei-plots}a maps $\rem$ as a function of magma-ocean thickness and forcing frequency for a dimensionless tidal potential magnitude of $\tidepotm=2\times10^{-8}$ that would be expected at about 4~Ga.

\subsection{Elliptical instability} \label{sec:Ei-instabilty}

To tentatively assess the propensity for elliptical instability of the tidally forced flow in the interior of the core, we adopt a widely used approach: we neglect viscosity to estimate the growth rate of perturbations, and compare this with the viscous damping rate of perturbations \citep{lacaze2004}. If growth exceeds damping, the system is considered unstable to perturbations in the form of inertial waves. Instability occurs by resonant growth of a pair of modes of the system that have an absolute frequency approximately half of the tidal frequency. These can be global eigenmodes but are often approximated as local, plane-wave inertial modes. These waves interact and break into turbulence that may, in turn, drive dynamo action.  With a BMO, the standard inertial eigenmodes will be complemented by interfacial--inertial modes, analogous to what appears in a stratified system \citep[e.g.,][]{kerswell1993elliptical}.

Analysis of the growth and damping rates of the elliptical instability in the core--BMO system is beyond the scope of this study. Instead, we apply the formalism developed for a single-layer system in an ellipsoidal geometry with no-slip boundary conditions  of \citet{grannan2016} and \citet{vidal2026did}. Here, the total growth rate $\sigma$ is the difference between inviscid growth and viscous damping,
\begin{equation}
    \label{eq:ei-growth-rate}
    \frac{\sigma}{\omega} \sim \frac{9}{16}\ellipticity - C_\delta(\ell)\Ek^{1/2} - C_\ellipticity(\ell)\Ek,
\end{equation}
where 
\begin{equation} \label{eq:Ek-def}
    \Ek \equiv \nu_C/\omega R_C^2
\end{equation}
is the Ekman number, of order $10^{-15}$ for Earth's core. We have assumed that the Moon's orbital frequency $n$ satisfies $n/\omega_\text{spin} \ll 1$ such that $\omega\sim\omega_\text{spin}$. The first term on the right-hand side, proportional to the ellipticity, is the inviscid growth rate; the second term represents viscous damping in a viscous layer at the no-slip boundary with the container; the third term represents viscous damping in the bulk interior. Factors $C_\delta$ and $C_\ellipticity$ depend on spherical harmonic degree; \citet{vidal2026did} find that the modes at $\ell \ge 7$ resonate and the dependencies on $\ell$ are well-captured by $C_\delta\propto \ell^{1/2}$ and $C_\ellipticity \propto \ell^3$, with order-one prefactors. 

For a very small Ekman number, the third term in \eqref{eq:ei-growth-rate} is negligible and viscous damping is dominantly at the boundary. Then the condition for instability ($\sigma\ge0$) becomes
\begin{equation} 
    \label{eq:ell-instability-number}
    \Ei \equiv \frac{R_C\ellipticity}{\sqrt{\nu_C/\omega}} \gtrsim \frac{C_\delta(\ell)}{9/16}\equiv \Ei_\text{crit},
\end{equation}
where we have defined $\Ei$ as a dimensionless number representing the ratio of inviscid growth and viscous (boundary-layer) damping, and $\Ei_\text{crit}$ as the threshold for instability. Note that $\Ei$ can also be written as $\Uel/\sqrt{\nu_C\omega}$ using the characteristic speed of the core's interior flow~\eqref{eq:elliptical-flow-speed}. It is plotted as a function of BMO thickness and forcing frequency in Figure~\ref{fig:rm-and-ei-plots}b.

Next we consider the critical value of $\Ei$, again emphasising that fundamentally distinct physics of the core--BMO system makes this discussion highly tentative. The elliptical instability threshold and finite-domain growth rate depend on global mode structure, boundary conditions and viscous damping \citep{vidal2026did}. These finite-domain effects are commonly encapsulated in the coefficient $C_\delta(\ell)$. In the present problem, the BMO modifies both the mode structure and the mechanical boundary conditions experienced by perturbations in the core, and therefore may alter $C_\delta(\ell)$ relative to single-layer, unstratified models (where modes are purely inertial). For example, tangential counterflow within the BMO may enhance viscous coupling and increase damping. Given the uncertainty, we adopt a representative value of $C_\delta \approx 2.8$ consistent with previous studies of the elliptical instability~\citep{kerswell1993elliptical, grannan2016, landeau2022sustaining}, in which case $\Ei_\text{crit} \approx 5$. Figure~\ref{fig:rm-and-ei-plots}b shows that $\Ei$ exceeds this threshold across a large region of parameter space. If instead $\Ei_\text{crit} \approx 50$, the threshold is exceeded only in a narrower region near the inviscid resonance from~\eqref{eq:inviscid-resonance-trend}. Further work is needed to constrain the growth, damping and mode selection of elliptical instability in the coupled core--BMO system --- particularly under conditions near the resonance identified here.

\subsection{Boundary-layer instability} \label{sec:BL-instability}

Our fully viscous tidal flow solution predicts a shear boundary layer in the core at the CMB. This tangential flow becomes strong at resonance. In some respects, its structure resembles that of a Stokes oscillating boundary layer \citep{Blondeaux_Vittori_2021}. Where such a boundary layer is driven by rotation, it is sometimes called an Ekman layer, and this terminology is common when considering precession or tidal flows \citep{legal2015flows, buffett2021conditions-7af}.  

In appendix~\ref{app:BL-structure}, we develop a rescaling of the problem in the limit of large (but finite) $\re$ to analyse the boundary-layer structure. We show that the overall core flow can be decomposed into an essentially inviscid interior flow driven by the tidal potential (\S\ref{sec:inviscid-solution}) and the oscillatory shear of a viscous boundary layer beneath the CMB. This shear flow has an amplitude that decays exponentially away from the CMB.  We show that the dimensional boundary-layer thickness $r_\delta \sim \sqrt{\nu_c/\omega}$. 

A similar flow is driven by in-plane oscillations of a flat plate. These induce an exponentially decaying velocity in the Stokes boundary layer of the adjacent fluid. As an appropriately defined Reynolds number is increased, such flows undergo, first, linear instability and, later, transition to turbulence. We use the notation of \cite{Blondeaux_Vittori_2021} and define the appropriate boundary-layer Reynolds number $R_\delta = U_\delta r_\delta/\nu_C$, where $U_\delta$ is the dimensional maximum zonal flow speed on the CMB in the equatorial plane.  Using the definition of the dimensionless zonal velocity given in equation~\eqref{eq:velocity-components-S-phi} and considering the non-dimensional scales in equation~\eqref{eq:characteristic-scales}, we obtain
\begin{equation}
    R_\delta 
    = \sqrt{\re/\Omega}\cdot  \max_{0\leq \phi<2\pi} \left| v_\phi (r=1,\,\theta=\pi/2,\,\phi) \right|. \label{eq:R_delta_def}
\end{equation}
Note that the factor $\sqrt{\re/\Omega}=\Ek^{-1/2} \omega^{-1}(4\pi G \rho_c)^{1/2}$, where the factor $\omega^{-1}(4\pi G \rho_c)^{1/2}\equiv \Omega^{-1}$ is associated with non-dimensionalising with respect to the gravitational timescale rather than the rotational time scale. Thus $R_\delta$ is proportional to $\Ek^{-1/2}$ and the boundary layer thickness is proportional to $\Ek^{1/2}$, consistent with \citet{legal2015flows}. 

Although oscillatory Ekman layers can radiate inertial waves into the fluid interior and generate internal shear layers \citep[e.g.,][]{greenspan1968TheoryRotatingFluids, kerswell1995internal}, the semidiurnal tidal frequency considered here lies close to the upper edge of the inertial-wave spectrum. We therefore expect boundary-layer excitation of inertial modes to be less likely to dominate than instability of the oscillatory shear layer itself.

\cite{Blondeaux_Vittori_2021} found that the critical condition for the Stokes boundary-layer instability to act over a significant part of the oscillatory cycle is $R_\delta \gtrsim 200$ and for it to act over the full cycle is $R_\delta \gtrsim 850$. Quantitatively similar thresholds govern transition to turbulence in Ekman layers \citep{buffett2021conditions-7af}. Figure~\ref{fig:rm-and-ei-plots}c shows that these thresholds are exceeded around the resonance peak and more broadly away from the resonance peak for a thin BMO. Thus instabilities of the oscillatory shear boundary layer near the CMB may contribute to the production of turbulence. The flow in our problem is more complex in several significant ways than the standard Stokes boundary-layer flow (spherical geometry, radial as well as tangential boundary motion, additional components of the flow). The flow is also more complex than that considered in studies of Ekman layers driven by tides or precession \citep{lorenzani2001fluid-23b}, because there is a coupled flow in the BMO and the maximum flow speed is enhanced. Moreover, shear flows of layered fluids may be susceptible to Kelvin-Helmholtz instabilities.

\section{Discussion}
\label{sec:discussion}

The theory developed above describes the laminar, tidally driven flow of the core--BMO system prior to the onset of inner-core solidification, neglecting Coriolis forces arising from rotation.  In this context, our results show that the inviscid solution within the core is formally identical to the canonical ellipsoidal vortex, and that the inviscid flow of the BMO responds to the cyclic rise and fall of the CMB.  Crucially, we discover a resonance when the forcing frequency matches an eigenmode of the system, i.e., when the tide advances at the same speed as an equatorial gravity wave with a wavelength that is half the circumference of the core.  In the inviscid case, this resonance frequency is a function of the thickness of the BMO and its density ratio with the core.  With finite viscous stress (in both fluids or just in the BMO), the resonance is damped. For a sufficiently small BMO-Reynolds number, viscous effects move the resonance trend to lower frequencies than the inviscid prediction. 

The viscosity of the basal magma ocean evolved with its thickness, temperature, composition and crystal fraction. Cooling would tend to increase viscosity, whereas compositional evolution of the liquid could offset or even reverse this trend \citep{Sun2020}.  Fractional crystallisation of bridgmanite would tend to enrich the residual liquid in iron, while the buoyancy of the crystallising assemblage and the flow dynamics would affect the entrainment or flotation of crystals \citep{Dragulet2024}. Even modest suspended crystal fractions substantially increase the effective viscosity of magma, and crystal-bearing magmas may exhibit non-Newtonian behaviour at sufficiently high crystal fraction or strain rate \citep{caricchi2007non}. Shear-driven crystal--liquid segregation could potentially introduce large-scale gradients in melt fraction and hence in effective viscosity \citep{Leighton1987}. These complications are probably most important late in BMO evolution, when the layer became thin, iron-rich and increasingly crystal-bearing. Given the uncertainty in the igneous petrology of the BMO, the present model treats viscosity as an independent control parameter rather than attempting to prescribe a unique rheological history.

\begin{figure}
    \centering
    \includegraphics[width=0.6\linewidth]{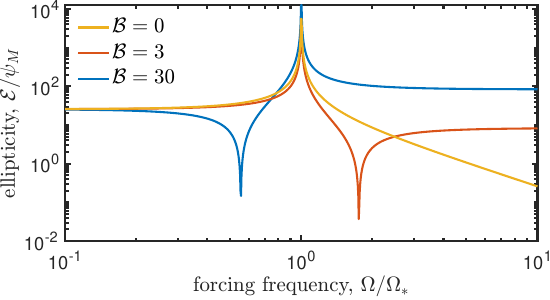}
    \caption{Forcing-normalised CMB ellipticity as a function of frequency for three values of solid-mantle ellipticity forcing $\curlyb$. In each case the flow is inviscid and other parameters are $\roc=1/2,\curlyr=1.05$. The resonance frequency $\Omega_*$ is independent of $\curlyb$, whereas the antiresonance frequency is a function of $\curlyb$. Physical consideration (Appendix~\ref{app:solid-mantle-forcing}) suggest $5 \lesssim \curlyb \lesssim 30$ over Earth history.}
    \label{fig:mantle-forcing-discussion}
\end{figure}

Previous studies of lunar tides of the core have not considered the BMO \citep[e.g.,][]{kerswell1994tidal, cebron2010systematic, vidal2026did}. In that case, and with the typical assumption of uniform density, the tidal potential acts on the core only by deforming the solid mantle. Hence previous work has forced the tidal flow by imposing a cyclic, ellipsoidal displacement of the CMB. With that boundary velocity as the sole forcing, its phase (relative to the Moon) is irrelevant. However, in our model, the primary forcing is the tidal gravitational potential.  If we incorporate the imposed motion of the solid mantle as an additional forcing, then the difference in phase of the two forcing mechanisms becomes important. For simplicity, our approach here is to fix the phase of the solid-mantle ellipsoid such that its long axis is aligned in the direction of the Moon. Details are provided in Appendix~\ref{app:solid-mantle-forcing} but the key outcome is shown in Figure~\ref{fig:mantle-forcing-discussion}. The combined forcing introduces an antiresonance in which forcing by the tidal potential and the solid mantle exactly cancel such that the ellipticity $\ellipticity$ is zero. The frequency of the (inviscid) antiresonance is a function of $\roc$, $\curlyr$ and $\mathcal{B}$. Fig.~\ref{fig:mantle-forcing-discussion} also shows that the solid-mantle forcing increases ellipticity at high frequency, but this is likely the consequence of neglecting the dynamics of solid-mantle tidal flexure.  A self-consistent approach would use the tidal potential as a forcing for the dynamics of the core, the BMO, and a viscoelastic solid mantle. The latter might function similarly to an ice shell over a water ocean \citep[e.g.,][]{matsuyama2018OceanTidal, rovira2023thin}. Furthermore, we have neglected spin-induced oblateness in the reference state, as is typical in tidal deformation models. Earth's oblateness today is a $\sim$0.06\% deviation-from-spherical, and a few percent deviation for the early Earth \citep[e.g.,][]{sabadini2016GlobalDynamics}. We leave these developments for future work.

The most significant limitation of the present theory is the omission of the Coriolis force. One concern is that rotation substantially alters the spectrum of free oscillations and could therefore suppress or shift the resonance with tidal forcing. Our preliminary rotating calculations do not support this outcome. Instead, the dominant resonance survives and undergoes only a modest frequency shift, whilst additional resonance branches emerge at higher forcing frequencies. This dominant mode arises from our generalisation of the degree-two eigenmode of a self-gravitating fluid sphere \citep[Kelvin's result;][]{lamb1924hydrodynamics}. It resonates when the semidiurnal tidal frequency, $\omega_\text{tide} = 2(\omega_\text{spin}-n)$, is approximately equal to the eigenfrequency. For $n\ll\omega_\text{spin}$, the tidal frequency lies close to the upper edge of the inertial spectrum, where inertial-wave motions become increasingly constrained. Inclusion of Coriolis therefore perturbs an existing interfacial gravity mode rather than replacing it. At the same time, restoring the Coriolis force couples spheroidal and toroidal motions, allowing a single degree-two, order-two forcing to excite a hierarchy of higher-degree interfacial gravity modes \citep[e.g.,][]{greenspan1964transient-b52,longuet-higgins1968EigenfunctionsLaplace, rieutord1997inertial}. The principal effect of rotation therefore appears to be a richer modal spectrum rather than elimination of the dominant resonant response. Consequently, we expect the resonance frequency, phase lag and viscous damping rate to differ quantitatively from the present theory, in which the Coriolis force is omitted. However, the qualitative behaviour is robust. It arises because the tidal potential acts on a density contrast separating two liquid layers, producing a resonant response through the restoring effects of self-gravity and fluid inertia. We therefore regard the present results as an analytically transparent limiting case that identifies the mechanism and its controlling parameters, rather than as a final quantitative prediction for the rotating Earth.

The advection of momentum is a nonlinearity that plays a central role in the cascade of tidal energy from the large wavelengths at which it is injected into the small wavelengths at which it is dissipated. Along the way, it sets up the dynamics of dynamo action and is hence of fundamental physical importance.  But it impedes mathematical analysis and hence we have set it aside in the present work.  However, as with the canonical ellipsoidal vortex model, our results provide a foundation onto which further analysis can be layered. In one such analysis, the nonlinear term provides the coupling that transmits energy from the tidally forced mode to eigenmodes of the unforced system \citep{kerswell2002elliptical}. Hence the tidally forced flow gives rise to elliptical instability, which represents a pathway to turbulence \citep{grannan2016, reun2016inertial-27e}.  In the context of our two-layer model, this pathway may be mediated by gravito-inertial modes. \cite{kerswell1993elliptical} studied such modes in an elliptically distorted cylindrical container with linear radial stratification.  He found that stratification enhances instability, particularly of the spinover mode.  The dynamics in a sphere or ellipsoid are more complicated and a fully numerical approach may be warranted \citep[e.g.,][]{favier2014non-linear-4e8, barker2016non-linear-b3c}.  An outstanding challenge for such models is to estimate level of turbulence at which nonlinear saturation caps the (resonant) kinetic energy \citep{vidal2026did}.

In light of the caveats above, further work is needed to refine the current theory and its predictions. The next step is to reintroduce the Coriolis force and re-derive the tidally driven flow---work that is in progress. We anticipate coupling of modes from degree-two, order-two to higher degrees and orders, complicating the mathematics but broadly preserving the behaviour, including the resonance \citep[e.g.,][]{hay2022HighFrequency, Pontin2023b, rovira2023thin}. Building on this, it will be necessary to reintroduce the nonlinear term and analyse the system through a variety of approaches, including analysis of parametric instabilities and full numerical simulation. Ultimately, a coupling with thermochemical convection, magnetic induction or consideration of the full magnetohydrodynamics may be warranted. Vigorous convection would give rise to an effective eddy viscosity that could mediate the transfer of tidal energy into the flow and, perhaps, the magnetic field \citep{Terquem21, deVries23, deVries23a, Terquem26}.

\subsection{Orbital--BMO history model}
\label{sec:orbital-history}

Despite the idealised state of our theory for core--BMO tides, an important consideration is whether its dynamics could potentially be relevant for the geodynamo of the early Earth, prior to the existence of the inner core. In particular, does the model support the hypothesis that core--BMO tides might have contributed to an early geodynamo?  Could the early Earth have accessed the resonance predicted here?

Of course the very existence of a BMO is an unproven prerequisite for this idea.  There are reasonable arguments that the BMO was inevitable \citep{boukare2025solidification, lark2026coupled}; here we take its existence for granted.  Furthermore, we assume that its thickness evolved within the range predicted by \cite{schaeffer2025energetically}.  To constrain the evolution of forcing frequency and magnitude, we use results by \cite{farhat2022resonant}. In that work, the Earth--Moon orbital distance and the length of day (LOD) are modelled consistently with observational constraints. We produce a simple, smooth fit to their more complicated models (details are in Appendix~\ref{app:trajectory}). We emphasise that prescribing the thermal--orbital evolution is not physically consistent because dissipation in the core--BMO system should feed back to modify it \citep[e.g.,][]{mignard1980EvolutionLunar, murray1999SolarSystem, Farhat2025}.  Future work should investigate this feedback.

\begin{figure}
    \centering
    \includegraphics [width=0.95\linewidth]{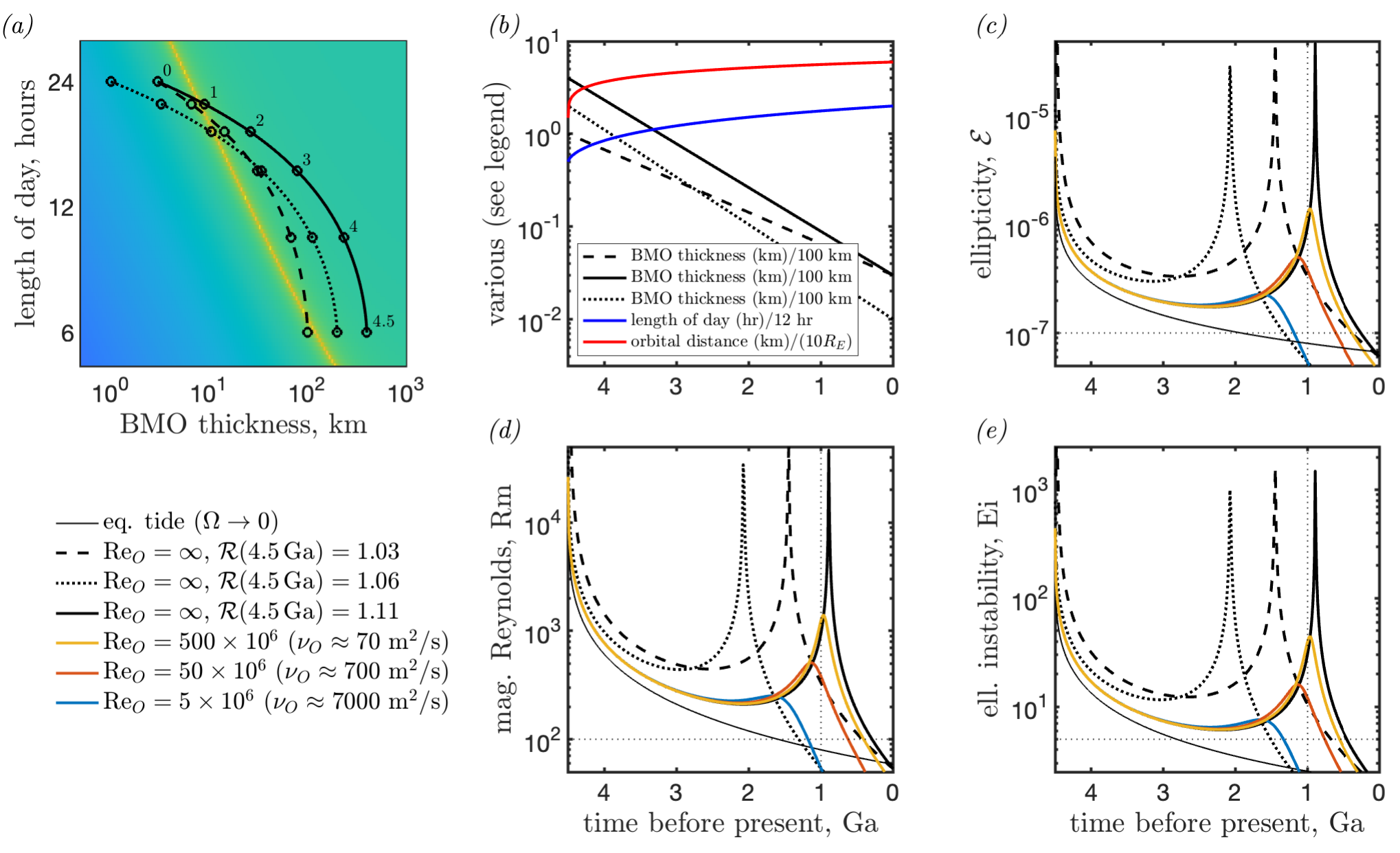}
    \caption{Hypothesised Earth-history trajectories and their consequences for the evolution of solution metrics. We consider three BMO thickness histories, starting at 100, 200 and 400~km and ending at 3, 1 and 3~km, respectively. Forcing frequency $\Omega(t)$ is cast in terms of Earth's length of day (LOD); see \S\ref{sec:tidal-forcing} above.  The evolution of LOD and lunar orbital distance are smooth, parameterised curves consistent with the detailed models of \cite{farhat2022resonant}. We assume an initial time of 4.5~Ga, after the Moon-forming impact. Orbital distance is used to compute $\tidepotm(t)$ by eq.~\eqref{eq:dimless-tidepot-definition}. All calculations use $\roc=1/2$. \textit{(a)} Colours indicate $\ellipticity/\tidepotm$ for the inviscid case (compare Fig.~\ref{fig:inviscid-maps}a), plotted as a function of LOD ($\propto 1/\Omega$) and dimensional BMO thickness ($\propto \curlyr-1$). Black curves show three trajectories with different BMO-thickness evolution, marked with time-points and their age in Ga. At 4.5~Ga, the BMO has a thickness of 400 (solid curve), 200 (dotted) or 100~km (dashed), spanning the range predicted by \cite{schaeffer2025energetically}. \textit{(b)} BMO thickness, LOD and orbital distance plotted against time before present. The decay constants for the three BMO thickness histories are given in Table~\ref{tab:orbital-evolution}. \textit{(c)} Ellipticity as a function of time in the past, accounting for $\tidepotm(t)$. The thin black curve is $\equil{\ellipticity} = 25.7\tidepotm$. Thick black curves are the inviscid solutions for the BMO scenarios. Thick coloured curves (see legend) represent different BMO viscosities in the ICVO formulation. \textit{(d)} Magnetic Reynolds number \eqref{eq:magnetic-reynolds} as a function of time before present.  Horizontal dotted line marks a value given as a possible minimum for dynamo action \citep{landeau2022sustaining}. Lines as in the legend. \textit{(e)} The elliptical instability number \eqref{eq:ell-instability-number} as a function of time before present. Horizontal dotted line marks a value given as a possible minimum for elliptical instability \citep{landeau2022sustaining}. }
    \label{fig:earth-history-model}
\end{figure}

In Figure~\ref{fig:earth-history-model}a,b, we hypothesise three simple, Earth-history `trajectories' representing BMO thickness, LOD and orbital distance from the Moon-forming impact $\sim$4.5~billion years ago (Ga) to the present. In terms of dimensionless variables, a trajectory is defined by $\curlyr(t), \,\Omega(t), \,\tidepotm(t)$. Our three trajectories, plotted as black lines in Fig.~\ref{fig:earth-history-model}a,b, are intended to be illustrative---the real trajectory is uncertain and likely more complicated. The trajectories differ in their initial BMO thickness (100, 200 and 400~km). They subsequently evolve by exponential decay to reach a small final thickness (1 or 3~km). Figure~\ref{fig:earth-history-model}a illustrates the variation of BMO thickness and LOD, overlaid on the inviscid ellipticity map. All three trajectories cross the inviscid resonance between about 0.9 and 2.1~Ga before present.  

Figure~\ref{fig:earth-history-model}c shows the evolution of inviscid ellipticity for the trajectories. For the case of the thicker BMO, curves representing the ICVO solution are plotted at three values of $\reo$, corresponding to dynamic viscosities in the range (4--400)$\times10^5$~Pa-s for $\rho_O=5000$~kg/m$^3$. A mantle-derived basaltic magma at 1200\textdegree{C} has a viscosity of about 100~Pa~s, so a realistic value of $\re_O$ could correspond to near-inviscid behaviour at resonance.  Note, however, that little is known about the evolution of BMO viscosity. Panel~(c) shows the equilibrium-tide ellipticity for comparison. This thin, black curve is equivalent to the ellipticity imposed in solid-mantle tidal models \citep{landeau2022sustaining} because it is linearly proportional to the magnitude of the tidal potential (recall that $\equil{\ellipticity}\approx25.7\tidepotm$). Using our orbital--BMO trajectories, we obtain $\equil{\ellipticity}\approx6\times10^{-8}$ at the CMB today (which is 60 percent of the standard value of $10^{-7}$; an acceptable discrepancy for present purposes). 

The magnetic Reynolds number $\rem$ and the elliptical instability number $\Ei$ are the most relevant metrics for assessing the potential for magnetic induction and mechanically driven turbulence. However, we emphasise that their supercriticality is a necessary but not sufficient condition for dynamo action and, furthermore, that their critical values are based on theory that neglects the BMO. With that caveat, the metrics are shown in Fig.~\ref{fig:earth-history-model}d,e. The evolution of $\rem$ and $\Ei$ closely follows that of ellipticity (panel~c), including the resonance peak that is tempered by viscous effects in the BMO. In each case, comparison of the dynamic solutions (thick lines) with the equilibrium-tide solution (thin line) shows that the presence of a BMO can substantially amplify both $\rem$ and $\Ei$. For the initially 100-km-thick BMO (dashed line), this amplification persists throughout most of the modelled history. The horizontal dotted lines mark heuristic critical values for magnetic induction and instability (\S\ref{sec:instabilities}). The dynamic solutions remain above these thresholds throughout the pre-inner-core era (up to $1\pm0.5$~Ga), although the excess above critical values is modest except near resonance. Moreover, the inviscid resonance peaks would likely be damped by turbulent and rotational effects that are not captured by the present theory. The palaeomagnetic record of field intensity over Earth history remains sparse, but \cite{Zhang2022} has reported evidence of a spike in intensity at $\sim$1.1~Ga (though these observations are debated by \cite{Huang2025}). This is intriguing but much more work would be needed to assess whether it is related to a tidal resonance.

The important point illustrated by Figure~\ref{fig:earth-history-model} is that a BMO can substantially enhance the efficiency with which lunar tides inject mechanical energy into the core. Whether this enhancement ultimately contributes to dynamo action depends on instability to turbulence in a rotating fluid (and its nonlinear saturation), magnetic induction, and/or interaction with convective flow, none of which are resolved here.

\subsection{Implications for thermal and orbital evolution}

The CMB ellipticity predicted by the model can be used to estimate an upper bound on the dimensional kinetic energy of the tidally forced core flow. We compute this estimate from equation~\eqref{eq:KE-inviscid-core}, using the time-dependent ellipticity $\ellipticity(t)$ and the energy scale $[E]$ defined in Table~\ref{tab:parameters}. In the inviscid histories shown in Figure~\ref{fig:earth-history-model}, $\ellipticity$ ranges from $\sim 10^{-7}$ to $\sim 10^{-5}$, depending on Earth--Moon distance, BMO thickness and proximity to resonance. The corresponding kinetic energies span approximately $10^{14}$--$10^{20}$~J. For comparison, \citet{landeau2022sustaining} estimate the kinetic energy of turbulent flow in the present-day outer core to be of order $10^{17}$~J. Thus, during near-resonant intervals, the large-scale tidal flow predicted here could have contained an amount of kinetic energy comparable to, or larger than, that associated with the present-day convective dynamo.

This comparison should be interpreted cautiously. The tidal response described by the present model is a coherent, large-scale flow, whereas dynamo action requires flow variations at smaller scale. The relevant quantity is therefore not the kinetic energy alone, but the fraction of that energy that is transferred to smaller scales by secondary instabilities and nonlinear interactions. The estimates of magnetic Reynolds number and elliptical-instability number in \S\ref{sec:instabilities} suggest that such transfer is plausible over parts of parameter space, especially near resonance, but the present linear theory cannot determine the efficiency of energy transfer to dynamo-relevant scales or the nonlinear saturation amplitude.

The viscous dissipation rates in the BMO shown in Figure~\ref{fig:dissipation-rate}b span dimensional values of approximately $0.01$--$10$~TW, with the largest values occurring near resonance. Dissipation in the core is typically about an order of magnitude smaller than dissipation in the BMO. These rates are comparable to the present-day terrestrial tidal dissipation, which is dominated by ocean tides and is approximately $3$~TW \citep{egbert2000SignificantDissipationa}. If comparable dissipation occurred in the deep past, BMO tides could have contributed to the rotational--orbital evolution of the Earth--Moon system by accelerating lunar recession and Earth's spin-down \citep{murray1999SolarSystem}. In the present calculation, however, the orbital history is prescribed rather than solved self-consistently. A coupled orbital--thermal model would be required to determine whether BMO dissipation significantly modifies the evolution of the Earth--Moon distance or the timing of resonance crossing \citep[e.g.,][]{downey2023ThermalOrbital}.

BMO dissipation could also contribute directly to the thermal evolution of the lowermost mantle and core. Present-day CMB heat flow is estimated to be of order $10$~TW \citep{lherm2024thermal}, comparable to the upper end of the dissipation rates predicted near resonance. Tidal dissipation within the BMO could therefore slow BMO crystallisation, alter the heat extracted from the core and feed back on the BMO thickness that controls the resonance itself. This feedback is not included in the prescribed BMO histories used in Figure~\ref{fig:earth-history-model}. Moreover, if the primary flow becomes unstable, turbulent dissipation may differ substantially from the laminar viscous dissipation calculated here, both in magnitude and in spatial distribution. The thermal and orbital implications of BMO tides therefore require a coupled model that includes tidal dissipation, BMO crystallisation, core cooling and lunar recession.

Although the model has been developed for Earth's early core, the mechanism is more general. Any differentiated body with an internal, two-layer structure of immiscible fluids of distinct densities (and with adequate tidal forcing) may support an analogous interfacial tidal response. The Moon may have hosted a BMO above its small liquid outer core \citep{scheinberg2018basal} and experienced episodes of enhanced tidal deformation during its orbital evolution \citep{touma1998ResonancesEarly, nimmo2024tidally-10e}. Mercury potentially crystallised from a magma ocean after a hit-and-run collision and has been affected by solar tides during capture into its 3:2 spin--orbit resonance \citep{soderlund2025puzzles}. Mars has a confirmed liquid core and evidence for a basal magma ocean \citep{lemaistre2023SpinState, samuel2023GeophysicalEvidence}. Although no large, tide-raising body orbits Mars today, one may have done so in the past \citep{arkani-hamed2009DidTidal}. Jupiter's moon Io, with its vigorous, tidally driven magmatism, has an iron core and possible evidence for an internally generated magnetic field \citep{khurana1997InteractionIo, kerswell1998TidalInstabilitya}. Ganymede provides a further example of a body in which a liquid core, stable stratification and dynamo action may interact \citep{christensen2015IronSnow}. Applying the present framework to these bodies would require only changes to the forcing potential, material parameters and orbital history, although rotation, magnetic induction and body-specific rheological models may be essential for quantitative predictions. More general still, but outside the scope of the current 2-layer configuration, are tides in stratified giant planets and stellar interiors where the general mechanism explored here is relevant \citep{Pontin2020, Pontin2023b, Pontin2023a, Horstmann2023}. 

\section{Conclusions}
\label{sec:conclusion}

We have developed a linear theory for lunar tides in a two-layer, early-Earth model consisting of a liquid-metal core overlain by a silicate basal magma ocean. In contrast to previous models that do not consider a BMO and in which the tide enters through an imposed deformation of a core--solid-mantle boundary, here the tidal potential acts directly on the core--BMO density contrast and dynamically deforms the internal interface.

The inviscid solution has two limiting behaviours. At low forcing frequency the response approaches an equilibrium tide set by the balance between tidal forcing and self-gravity. At higher forcing frequency the core--BMO interface supports an interfacial gravity wave, and the forced response becomes resonant when the tidal frequency matches the natural frequency of the degree-two interfacial mode. The resulting core flow ($r<1$) is formally identical to the canonical elliptical vortex, providing a direct connection between the present two-layer tidal problem and the classical theory of elliptical instability.

Finite viscosity regularises the inviscid resonance, introduces a phase lag between the tidal potential and the CMB ellipticity, and produces oscillatory boundary layers at the CMB and at the top of the BMO. The viscous response is controlled primarily by BMO thickness and viscosity. A BMO that is viscously dominated (with a small effective Reynolds number $\re_\text{BMO}$) dampens the resonance, reduces ellipticity, and suppresses the associated core-flow speed.

When combined with simple histories for lunar recession, Earth's spin-down and BMO thinning, the model predicts that the early Earth could have crossed a resonant regime in which the BMO amplified tidally driven core motion. Across plausible parameter ranges, the resulting magnetic Reynolds number and elliptical-instability number exceed commonly used critical values, demonstrating that a BMO can substantially amplify tidal coupling to the core and suggesting that the resulting flow may have contributed to mechanically driven turbulence and magnetic induction.

These results should be interpreted as an analytically transparent limiting case. We have neglected the Coriolis force, nonlinear advection of momentum, self-consistent deformation of the solid mantle and magnetic coupling. We expect that restoring the Coriolis force will modify the modal structure and quantitative resonance conditions, but preserve the central mechanism: a deformable density interface that allows the tidal potential to excite a resonant response that is absent from single-layer, imposed-ellipticity models. Future work should incorporate rotation, nonlinear instability, coupling to the magnetic field, and coupled thermal--orbital evolution to assess whether this mechanism made a significant contribution to Earth's early magnetic field.


\section*{Acknowledgements} The authors thank A.~Barker, D.~C\'{e}bron, R.~Kerswell, S.~Labrosse, M.~Landeau, F.~Nimmo, and N.~Schaeffer for discussion and suggestions.  We thank M.~Kameyama and two anonymous reviewers for comments that helped us to improve the manuscript.

\section*{Funding} 
RK and HH were supported by Leverhulme Project Grant RPG-2021-199. J.F.J.B. was supported by the UKRI Research Frontier Guarantee program EP/Y014375/1. MK was supported by a Research Experience Placement grant from NERC and by the St Andrews Research Internship Scheme (StARIS).

\section*{Declaration of interests}
The authors report no conflict of interest.

\section*{Data availability statement}
The data (code) that support the findings of this study are openly available in \cite{code-repo} at \url{https://doi.org/10.5281/zenodo.15492234}.

\section*{Author ORCiDs} 
J.F.J.~Bryson \href{https://orcid.org/0000-0002-5675-8545}{0000-0002-5675-8545},
H.C.F.C.~Hay \href{https://orcid.org/0000-0003-1746-1228}{0000-0003-1746-1228},
R.F.~Katz \href{https://orcid.org/0000-0001-8746-5430}{0000-0001-8746-5430}, 
M.B.C.~Kiernan \href{https://orcid.org/0009-0006-2476-1084}{0009-0006-2476-1084},
D.W.~Rees~Jones \href{https://orcid.org/0000-0001-8698-401X}{0000-0001-8698-401X}.

\section*{Author contributions}
Katz conceived the hypothesis in discussion with Bryson. Katz and Hay formulated the mathematical problem. Kiernan solved the problem, supervised by Katz and Hay. Rees~Jones, Hay, Katz and Kiernan analysed the solution. Katz produced the figures and led in drafting the manuscript.  All authors contributed to drafting and editing the manuscript.

\section*{Use of artificial intelligence}
Artificial intelligence has been used in checking code for errors and making minor changes to human-written code.

\appendix

\section{Boundary conditions}
\label{app:BCs}

In this appendix we develop the system of linear equations that represents the application of the boundary conditions to constrain the solution coefficients $A\elem,B\elem,C\elem,D\elem,E\elem,F\elem$ and $r\elem$.  We consider the problem with a viscous core and viscous BMO, and hence begin with the boundary conditions \eqref{eq:boundary-conditions-linearised}, which constrain both the velocity and viscous stress.  Our first objective is therefore to express velocity and stress in terms of the unknown coefficients.

We begin with an expression for the perturbation stress tensor $\pert{\stressten} = -\pert{P}\identity + {\devstressten}$, and hence the deviatoric stress tensor ${\devstressten}$.  This is a dimensionless perturbation quantity given by
\begin{equation}
    {\devstressten} = \frac{\varrho_h\nu_h}{\re}\left[\Grad\vel + (\Grad\vel)^T\right]\equiv \frac{1}{2}\lmsumh\tidepot\elem\devstressten\elem + \cc,
\end{equation}
where the equivalence provides a definition for $\devstressten\elem$. With this and the velocity solution \eqref{eq:velocity-components-S}, the components of the traction vector on a sphere are
\begin{subequations}
    \label{eq:stress-components}
    \begin{align}
        \rhat\cdot {\devstressten}\elem\cdot\rhat &= 2\ell(\ell+1)\frac{\varrho_h\nu_h}{\re}\left[-\frac{2}{r^3}{S}\elem + \frac{1}{r^2}\fd{{S}\elem}{r}\right]\sph{\ell}{m},\\
        \rhat\cdot{\devstressten}\elem \cdot\that &= \frac{\varrho_h\nu_h}{\re}\left[\frac{\ell(\ell+1)}{r^3}{S}\elem - \frac{2}{r^2}\fd{{S}\elem}{r} + \frac{1}{r}\fd{^2{S}\elem}{r^2}\right]\pd{\sph{\ell}{m}}{\theta},\\
        \rhat\cdot{\devstressten}\elem \cdot\phat &= \frac{\varrho_h\nu_h}{\re}\left[\frac{\ell(\ell+1)}{r^3\sin\theta}{S}\elem - \frac{2}{r^2\sin\theta}\fd{{S}\elem}{r} + \frac{1}{r\sin\theta}\fd{^2{S}\elem}{r^2}\right]\pd{\sph{\ell}{m}}{\phi}.
    \end{align}
\end{subequations}
Now substituting the solution for $S\elem$ from \eqref{eq:solution-S} into these expressions, and also into \eqref{eq:velocity-components-S} for the velocity, and using Bessel-function identities,
\begin{subequations}
\begin{align}
    {\vel\elem}\cdot\rhat =  \sph{\ell}{m}&\begin{cases}
        d^{\ell m}_rD\elem+a^{\ell m}_{r}A\elem & r<1,\\
        e^{ \ell m}_{r}E\elem+f^{\ell m}_{r}F\elem+b^{\ell m}_{r}B\elem+c^{\ell m}_{r}C\elem & 1<r<\curlyr,
    \end{cases}\\
    {\vel\elem}\cdot\that =  \pd{\sph{\ell}{m}}{\theta}&\begin{cases}
        d^{\ell m}_{t}D\elem+a^{\ell m}_{t}A\elem & r<1,\\
        e^{\ell m}_{t}E\elem+f^{\ell m}_{t}F\elem+b^{\ell m}_{t}B\elem+c^{\ell m}_{t}C\elem & 1<r<\curlyr,
    \end{cases}\\
    {\vel\elem}\cdot\phat =  \frac{1}{\sin\theta}\pd{\sph{\ell}{m}}{\phi}&\begin{cases}
        d^{\ell m}_{t}D\elem+a^{\ell m}_{t}A\elem & r<1,\\
        e^{\ell m}_{t}E\elem+f^{\ell m}_{t}F\elem+b^{\ell m}_{t}B\elem+c^{\ell m}_{t}C\elem & 1<r<\curlyr,
    \end{cases}\\
    \rhat\cdot{\devstressten\elem} \cdot\rhat =  \sph{\ell}{m}&\begin{cases}
      \delta^{\ell m}_{r}D\elem + \alpha^{\ell m}_{r}A\elem & r<1,\\
      \epsilon^{\ell m}_{r}E\elem+\zeta^{\ell m}_{r}F\elem+\beta^{\ell m}_{r}B\elem+\gamma^{\ell m}_{r}C\elem & 1<r<\curlyr,
    \end{cases}\\
    \rhat\cdot{\devstressten\elem} \cdot\that =  \pd{\sph{\ell}{m}}{\theta}&\begin{cases}
        \delta^{\ell m}_{t}D\elem+\alpha^{\ell m}_{t}A\elem & r<1,\\
        \epsilon^{lm}_{t}E\elem+ \zeta^{\ell m}_{t}F\elem+\beta^{\ell m}_{t}B\elem+\gamma^{\ell m}_{t}C\elem & 1<r<\curlyr,
        \end{cases}\\
     \rhat\cdot{\devstressten\elem} \cdot\phat =  \frac{1}{\sin\theta}\pd{\sph{\ell}{m}}{\phi}&\begin{cases}
        \delta^{\ell m}_{t}D\elem+\alpha^{\ell m}_{t}A\elem & r<1,\\
        \epsilon^{lm}_{t}E\elem+ \zeta^{\ell m}_{t}F\elem+\beta^{\ell m}_{t}B\elem+\gamma^{\ell m}_{t}C\elem & 1<r<\curlyr
    \end{cases}
\end{align}
\end{subequations}
where we have defined $\vel\elem$ such that $\vel\equiv\tfrac{1}{2}\lmsumh\tidepot\elem\vel\elem + \cc.$ Here, for concision in writing the boundary conditions below, we have introduced a series of radial functions for the velocities,
\begin{equation} \label{eq:app-BCs-velocities}
\begin{aligned} 
    a^{\ell m}_r (r) &= \frac{\ell(\ell+1)}{im\Omega}r^{\ell-1} = b^{\ell m}_r(r),\\
    c^{\ell m}_r(r) &= \frac{\ell(\ell+1)}{im\Omega}r^{-\ell-2},\\
    d^{\ell m}_r(r) &= \ell(\ell+1)r^{-\frac{3}{2}}J_{\ell+\frac{1}{2}}(q^{m}_Cr),\\
    e^{\ell m}_r(r) &= \ell(\ell+1)r^{-\frac{3}{2}}J_{\ell+\frac{1}{2}}(q^{m}_Or),\\
    f^{\ell m}_r(r) &= \ell(\ell+1)r^{-\frac{3}{2}}J_{-\ell-\frac{1}{2}}(q^{m}_Or)
\end{aligned}\qquad\quad\begin{aligned}
    a^{\ell m}_t (r) &= \frac{\ell+1}{im\Omega}r^{\ell-1} = b^{\ell m}_{t}(r),\\
    c^{\ell m}_{t}(r) &= -\frac{\ell}{im\Omega}r^{-\ell-2},\\
    d^{\ell m}_t(r) &= (\ell+1)r^{-\frac{3}{2}}J_{\ell+\frac{1}{2}}(q^{m}_Cr)-q^{m}_Cr^{-\frac{1}{2}}J_{\ell+\frac{3}{2}}(q^{m}_Cr),\\
    e^{\ell m}_{t}(r) &= (\ell+1)r^{-\frac{3}{2}}J_{\ell+\frac{1}{2}}(q^{m}_Or)-q^{m}_Or^{-\frac{1}{2}}J_{\ell+\frac{3}{2}}(q^{m}_Or),\\
    f^{\ell m}_{t}(r) &= (\ell+1)r^{-\frac{3}{2}}J_{-\ell-\frac{1}{2}}(q^{m}_Or)+q^{m}_Or^{-\frac{1}{2}}J_{-\ell-\frac{3}{2}}(q^{m}_Or),
\end{aligned}
\end{equation}
and for the stresses,
\begin{subequations}
\begin{equation} \label{eq:app-BCs-stresses-A}
\begin{aligned}    
    \alpha^{\ell m}_r(r) &= 2\frac{\ell(\ell^2-1)}{im\Omega\re}r^{\ell-2},\\
    \beta^{\ell m}_r(r) &= 2\frac{\ell(\ell^2-1)\roc\noc}{im\Omega\re}r^{\ell-2},\\
    \gamma^{\ell m}_r(r) &= -2\frac{\ell(\ell+1)(\ell+2)\roc\noc}{im\Omega\re}r^{-\ell-3},
    \end{aligned} \qquad \qquad \begin{aligned}
     \alpha^{\ell m}_{t}(r) &= 2\frac{\ell^2-1}{im\Omega\re}r^{\ell-2},\\
     \beta^{\ell m}_{t}(r) &= 2\frac{(\ell^2-1)\roc\noc}{im\Omega\re}r^{\ell-2},\\
    \gamma^{\ell m}_{t}(r) &= 2\frac{\ell(\ell+2)\roc\noc}{im\Omega\re}r^{-\ell-3},
    \end{aligned} 
    \end{equation}
    \begin{equation} \label{eq:app-BCs-stresses-B}
    \begin{aligned}
    \delta^{\ell m}_r(r) &= 2\frac{\ell(\ell+1)}{\re}\left[(\ell-1)r^{-\frac{5}{2}}J_{\ell+\frac{1}{2}}(q^{m}_Cr)-q^{m}_Cr^{-\frac{3}{2}}J_{\ell+\frac{3}{2}}(q^{m}_Cr)\right],\\
    \epsilon^{\ell m}_r(r) &= 2\frac{\ell(\ell+1)\roc\noc}{\re}\left[(\ell-1)r^{-\frac{5}{2}}J_{\ell+\frac{1}{2}}(q^{m}_Or)-q^{m}_Or^{-\frac{3}{2}}J_{\ell+\frac{3}{2}}(q^{m}_Or)\right],\\
    \zeta^{\ell m}_r(r) &= 2\frac{\ell(\ell+1)\roc\noc}{\re}\left[(\ell-1)r^{-\frac{5}{2}}J_{-\ell-\frac{1}{2}}(q^{m}_Or)+q^{m}_Or^{-\frac{3}{2}}J_{-\ell-\frac{3}{2}}(q^{m}_Or)\right],\\ 
    \delta^{\ell m}_{t}(r) &= \frac{1}{\re}\left\{\left[2(\ell^2-1)-(q^{m}_C)^2r^2\right]r^{-\frac{5}{2}}J_{\ell+\frac{1}{2}}(q^{m}_Cr)+2q^{m}_Cr^{-\frac{3}{2}}J_{\ell+\frac{3}{2}}(q^{m}_Cr)\right\},\\
    \epsilon^{\ell m}_{t}(r) &= \frac{\roc\noc}{\re}\left\{\left[2(\ell^2-1)-(q^{m}_O)^2r^2\right]r^{-\frac{5}{2}}J_{\ell+\frac{1}{2}}(q^{m}_Or)+2q^{m}_Or^{-\frac{3}{2}}J_{\ell+\frac{3}{2}}(q^{m}_Or)\right\},\\
    \zeta^{\ell m}_{t}(r) &= \frac{\roc\noc}{\re}\left\{\left[2(\ell^2-1)-(q^{m}_O)^2r^2\right]r^{-\frac{5}{2}}J_{-\ell-\frac{1}{2}}(q^{m}_Or)-2q^{m}_Or^{-\frac{3}{2}}J_{-\ell-\frac{3}{2}}(q^{m}_Or)\right\},
\end{aligned}
\end{equation}
\end{subequations}
where we recall that $J_n$ is a Bessel function of the first kind, of order $n$.

We are finally in a position to formulate the linear system of equations, corresponding to the boundary conditions \eqref{eq:boundary-conditions-linearised}, that constrain the set of unknown coefficients.  This system is
\begin{equation}
    \label{eq:boundary-condition-linear-system}
    \begin{bmatrix}
        0 & b^{\ell m}_r(\curlyr) & c^{\ell m}_r(\curlyr) & 0 & e^{\ell m}_r(\curlyr) & f^{\ell m}_r(\curlyr) & 0 \\
        0 & b^{\ell m}_t(\curlyr) & c^{\ell m}_t(\curlyr) & 0 & e^{\ell m}_t(\curlyr) & f^{\ell m}_t(\curlyr) & 0 \\
        a^{\ell m}_r(1) & -b^{\ell m}_r(1) & -c^{\ell m}_r(1) & d^{\ell m}_r(1) & -e^{\ell m}_r(1) & -f^{\ell m}_r(1) & 0 \\
        a^{\ell m}_t(1) & -b^{\ell m}_t(1) & -c^{\ell m}_t(1) & d^{\ell m}_t(1) & -e^{\ell m}_t(1) & -f^{\ell m}_t(1) & 0 \\
        M_{51} & M_{52} & M_{53} & \delta^{\ell m}_r & -\epsilon^{\ell m}_r & -\zeta^{\ell m}_r & M_{57} \\
        \alpha^{\ell m}_t(1) & -\beta^{\ell m}_t(1) & -\gamma^{\ell m}_t(1) & \delta^{\ell m}_t(1) & -\epsilon^{\ell m}_t(1) & -\zeta^{\ell m}_t(1) & 0 \\
        a^{\ell m}_r(1) & 0 & 0 & d^{\ell m}_r(1) & 0 & 0 & im\Omega \\        
    \end{bmatrix}
    \begin{bmatrix}
        A\elem \\ B\elem \\ C\elem \\ D\elem \\ E\elem \\ F\elem \\ r\elem
    \end{bmatrix} = 
    \begin{bmatrix}
       Q\delta_{\ell 2}\delta_{m2} \\ 0 \\ 0 \\ 0 \\ 1-\roc \\ 0 \\ 0
    \end{bmatrix}
\end{equation}
where $Q = -i\Omega\curlyr\curlyb$ and 
\begin{displaymath}
    \begin{aligned}
        M_{51} &= -(\ell+1) + \alpha_r^{\ell m}(1), \\
        M_{52} &= (\ell+1)\roc - \beta_r^{\ell m}(1), \\
        M_{53} &= -\ell\roc - \gamma_r^{\ell m}(1), \\
        M_{57} &= (1-\roc)\left(\frac{1}{3} - \frac{1-\roc}{2\ell+1}\right).
    \end{aligned}
\end{displaymath}
In equation~\eqref{eq:boundary-condition-linear-system}, the ordering of rows follows the conditions \eqref{eq:boundary-conditions-linearised}; the first pair of rows is (modified) no-slip at $r=\curlyr$,  the second pair is continuity of velocity across $r=1$, the third pair is continuity of stress across $r=1$, and the last row is the kinematic condition at $r=1$.  Each pair comprises a radial and a tangential component.

\section{Boundary-layer structure analysis} \label{app:BL-structure}
In this appendix, we analyse the structure of the boundary-layer at the core--magma boundary (CMB) by rescaling the boundary condition matrix from appendix~\ref{app:BCs}.

The coefficients in equations~(\ref{eq:app-BCs-velocities}, \ref{eq:app-BCs-stresses-B}) in the boundary-condition matrix~\eqref{eq:boundary-condition-linear-system} involve various Bessel functions of the form $J_\nu(z)$ where $z=q_C^m r$ or $z=q_O^m r$. For $\re\rightarrow \infty$, $|z| \sim (\Omega \re)^{1/2}$ (up to a prefactor), and $\arg z=\pm \pi/4$, where the sign corresponds to the sign of $m$. The standard properties of Bessel functions mean that their magnitude increases exponentially with $\re$. To account for this behaviour, we work in terms of a scaled Bessel function
$$ \tilde{J}_\nu(z) = J_\nu(z) \exp\left\{-|\Im (z)|\right\},$$
where $\Im(z)$ denotes the imaginary part of $z$. The scaled Bessel functions $\tilde{J}_\nu$ still depend on $z$, but this rescaling captures the dominant exponential dependence.
In particular, the coefficients $d^{\ell m}_r$, $d^{\ell m}_t$, $\delta^{\ell m}_r$, $\delta^{\ell m}_t$ are all proportional to terms of the form $J_{\nu}(q^{m}_Cr)$. But $|\Im (q_C^m r) | = r (|m|/2)^{1/2}(\Omega\re)^{1/2} $, and so 
$$ \tilde{J}_\nu(q^{m}_Cr) = J_\nu(q^{m}_Cr) \exp\left\{-r (|m|/2)^{1/2}(\Omega\re)^{1/2}\right\}.$$
All these coefficients appear in the boundary-condition-matrix equations evaluated at $r=1$ and multiplying $D_\ell^{m}$. This motivates a rescaling 
$$\tilde{D}_\ell^{m} = D_\ell^{m} \exp\left\{(|m|/2)^{1/2}(\Omega\re)^{1/2}\right\}.$$

A similar rescaling can be applied with all the coefficients $e^{\ell m}_r$, $e^{\ell m}_t$, $\epsilon^{\ell m}_r$, $\epsilon^{\ell m}_t$ which appear multiplying $E_\ell^{m}$, and also with the coefficients $f^{\ell m}_r$, $f^{\ell m}_t$, $\zeta^{\ell m}_r$, $\zeta^{\ell m}_t$ which appear multiplying $F_\ell^{m}$. In these cases, $\Omega\re$ is replaced by $\Omega\re/{\nu_{O/C}}$ from the definition of $q^m_O$, and the coefficients are evaluated at $r=\curlyr$ in the matrix~\eqref{eq:boundary-condition-linear-system}. 

In order to evaluate the structure function of the velocity field $S\elem$ in the core ($r<1$) using equation~\eqref{eq:solution-S}, we note that there are two contributions, one involving $D\elem$ and the other involving $A\elem$. The first of these contributions, denoted $S_1$ can be rewritten
\begin{align}
     S_1&\equiv r^\frac{1}{2} D\elem J_{\ell+\frac{1}{2}}(q^m_Cr),  \nonumber  \\
    &=r^\frac{1}{2}  \tilde{D}\elem \exp\left\{(-|m|/2)^{1/2}(\Omega\re)^{1/2}\right\} \tilde{J}_{\ell+\frac{1}{2}}(q^m_Cr) \exp\left\{+r (|m|/2)^{1/2}(\Omega\re)^{1/2}\right\},  \nonumber  \\
    &=r^\frac{1}{2}  \tilde{D}\elem \tilde{J}_{\ell+\frac{1}{2}}(q^m_Cr) \exp\left\{-(1-r)/\tilde{r}_\delta \right\}, \label{eq:pert_S1}
\end{align}
where
\begin{equation} 
    \label{eq:CMB_boundary_thickness_scale}
    \tilde{r}_\delta \equiv (|m|/2)^{-1/2}(\Omega\re)^{-1/2},
\end{equation}
is a dimensionless boundary-layer thickness scale. The dimensional equivalent scale given in \S\ref{sec:BL-instability} can be obtained by setting $m=2$ and using the scales in equation~\eqref{eq:characteristic-scales}. Therefore, the first contribution to the structure function near $r=1$ represents a viscous boundary-layer flow driven by the oscillating motion of the CMB, the thickness of which decreases as the Reynolds number increases. 

The second contribution, denoted $S_2$, 
\begin{equation} 
    \label{eq:pert_S2}
    S_2\equiv \frac{A\elem}{im\Omega}r^{\ell+1}
\end{equation}
is a flow driven by the tidal gravitational body force.

While the above rescaling successfully captures the dominant (exponential) dependence on the Reynolds number and uncovers the radial structure of the boundary layer, the resulting boundary condition matrix retains some (algebraic) dependence on Reynolds number and the other parameters of the system. This dependence gives rise to the non-resonant and resonant behaviour shown in Fig.~\ref{fig:viscous-Re-scaling} (panels a and b respectively).  

In general, both contributions play a role in $S\elem$ and hence to the radial flow. Due to the exponential decay envelope in equation~\eqref{eq:pert_S1} away from $r=1$, $S_1$ decays rapidly, leaving $S_2$ to dominate away from the CMB boundary-layer flow. Thus the radial part of core flow is driven dominantly by the tidal gravitational body force. 

The viscous boundary-layer flow characterized by $S_1$ has an amplitude that decays approximately exponentially away from $r=1$. For sufficiently large $\re$, we can replace the weak $r$-dependence in equation~\eqref{eq:pert_S1} by the approximation 
\begin{equation}
    S_1 \approx  \tilde{D}\elem \tilde{J}_{\ell+\frac{1}{2}}(q^m_C) \exp\left\{-\frac{1-r}{\tilde{r}_\delta} \right\}
\end{equation}
This induces a component of the longitudinal velocity proportional to 
\begin{equation}
    \frac{d S_1}{dr} \approx  \frac{\tilde{D}\elem \tilde{J}_{\ell+\frac{1}{2}}(q^m_C)}{\tilde{r}_\delta} \exp\left\{-\frac{1-r}{\tilde{r}_\delta} \right\} \propto \exp\left\{-\frac{1-r}{\tilde{r}_\delta} \right\},
\end{equation}
where the constant of proportionality will depend on $\Omega$ and the other parameters of the system.   
In turn, the shear-strain rate $\partial_r{v}_\phi$ is proportional to $\left|\infd^2 S_1/\infd r^2\right|$. The additional derivative introduces an additional factor of ${\tilde{r}_\delta}^{-1} \propto \re^{1/2}$ from equation~\eqref{eq:CMB_boundary_thickness_scale}, consistent with the calculations shown in Fig.~\ref{fig:viscous-Re-scaling}.

\section{Potential, kinetic and dissipation rate of energy}
\label{app:energy}

\textbf{Potential energy.} At any point within the domain, the dimensional potential energy per unit volume in excess of the hydrostatic base state is given by $e_P(\rpos) = [\rho(\rpos)-\rho_h(\rpos)]g(\rpos)(r-R_C)$, where $\rho_h$ represents the hydrostatic density structure. The volume of fluid between the tidally displaced CMB and its reference position is denoted $V_P$ and is the region where $e_P$ is non-zero. Since the CMB displacement is formally small, we linearise by defining a local coordinate $\pert{r} = r - R_C \ll 1$ and taking $g(\rpos)\sim g_h(R_C)$. Hence, in terms of dimensionless symbols,
\begin{align*}
    E_P &= \int_{V_P}e_P \,\infd V,\nonumber\\ &\sim (1-\roc)g_h(1)\int_0^{2\pi}\infd\phi\int_0^\pi\infd\theta\int_0^{\pert{r}_C(\theta,\phi)}\infd \pert{r}\, |\pert{r}|\sin\theta,
\end{align*}
where the total potential energy $E_P$ is made dimensionless with $[E] = 4\pi G\rho_C^2R_C^5$. Substituting \eqref{eq:sphere-harm-ansatz-rC} for $\pert{r}_C$, using $g_h(1) = 1/3$ and integrating, while making use of the unnormalised spherical harmonic orthogonality conditions, gives
\begin{equation}
    \label{eq:potential-energy}
    E_P \sim (1-\roc)\frac{\tidepotm^2}{4}\lmsumh N\elem|r\elem|^2,
\end{equation}
where all symbols are dimensionless and the normalisation factor is 
\begin{equation}
    \label{eq:normalisation-factor}
    N\elem \equiv \frac{4\pi}{2\ell+1}\frac{(\ell+m)!}{(\ell-m)!},\quad\qquad(\text{i.e., }N_2^2\approx60.3).
\end{equation}
Inserting \eqref{eq:ellipticity-rc} and \eqref{eq:dimless-tidepot-definition} into \eqref{eq:potential-energy} gives the dimensionless potential energy of the inviscid solution for the $\ell=m=2$ mode as equation~\eqref{eq:EP-inviscid}. \\

\noindent\textbf{Kinetic energy.} At any point within the domain, the dimensional kinetic energy per unit volume is given by $e_K(\rpos) = \rho\vel\cdot\vel/2$. The total kinetic energy within a radius $r_i \le r \le r_o$ is given, in dimensionless terms, by
\begin{align}
    E_K(r_i,r_o) &= \frac{1}{2}\int_0^{2\pi}\infd\phi\int_0^\pi\infd\theta\int^{r_o}_{r_i}\infd r \rho(\vel\cdot\vel)r^2\sin\theta, \nonumber \\
    &\equiv \int^{r_o}_{r_i} r^2\solidangle{e}_K \, \infd r, 
    \label{eq:kinetic-energy-radial-integral}
\end{align}
where $\solidangle{e}_K$ is the radial kinetic-energy density. Again making use of the spherical harmonic orthogonality relations, the above integral over $4\pi$ solid angle gives
\begin{equation}
    \label{eq:KE-solid-angle}
    \solidangle{e}_K(r) = \rho\lmsumh \left(\tidepot\elem\right)^2\ell(\ell+1)\frac{N\elem}{2r^2}\left(\ell(\ell+1)\frac{|S\elem|^2}{r^2} + \left\vert\fd{S\elem}{r}\right\vert^2\right).
\end{equation}
With this we can write $E_K(\text{core}) = \int_0^1 r^2 \solidangle{e}_K\,\infd r$ and $E_K(\text{BMO}) = \int_1^\curlyr r^2 \solidangle{e}_K\,\infd r$. The kinetic energy of the core in the inviscid solution is obtained by combining \eqref{eq:inviscid-analytical-coefs} with \eqref{eq:solution-S} for $S\elem$. For the $2,2$ inviscid mode, the total dimensionless kinetic energy of the core is given by equation~\eqref{eq:KE-inviscid-core}. \\

\noindent\textbf{Dissipation rate.} The dimensional dissipated energy density is $\dot{e}_\nu (\rpos) = 2\rho \nu \strainrten : \strainrten$, where the strain-rate tensor is $\strainrten = \tfrac{1}{2}\left[\Grad{\pert\vel} + \left(\Grad{\pert\vel} \right)^T \right]$. Rescaling with $[E]/[t]$, the total dimensionless viscous dissipation rate within a volume bounded by $r_i \leq r \leq r_o$ is 
\begin{align}
    \dot{E}_\nu (r_i, r_o) &= \frac{2}{\re}\int_0^{2\pi} \infd\phi \int_0^\pi \infd\theta \int_{r_i}^{r_o} \infd r \left(\varrho_h\nu_h \, \strainrten : \strainrten \, r^2 \sin\theta\right), \\
    &\equiv \int_{r_i}^{r_o} r^2 \solidangle{e}_\nu \infd r
\end{align}
where $\solidangle{e}_\nu$ is the dimensionless radial viscous dissipation-rate density, and we note that
\begin{equation}
    \varrho_h\nu_h = \begin{cases}
        1 & r \le 1, \\
        \roc \noc & 1 < r < \curlyr.
    \end{cases}
\end{equation}
The dimensionless viscous dissipation rate per radial distance is derived using the vector spherical harmonic orthogonality conditions in A.19 and A.20 of \citet{rieutord1987linear}, giving
\begin{align}
    \label{eq:diss_viscous_angavg}
    \solidangle{e}_\nu &\equiv \int_0^{2\pi} \infd \phi \int_0^\pi \infd \theta \left(\dot{e}_{\nu} \sin \theta \right),\notag \\  &=  \tidepotm^2\frac{\varrho_h\nu_h}{2\re}\lmsumh\frac{N\elem}{r^2} \Bigg(
    2 r^2 \left|\frac{\infd \pert{u}\elem}{\infd r}\right|^2 + \left|2\pert{u}\elem - \ell(\ell+1)\pert{v}\elem \right|^2 +
    \ell(\ell+1) \left| \pert{u}\elem -  \pert{v}\elem + 
    r\frac{\infd \pert{v}\elem}{\infd r}  \right|^2  \notag \\ 
    &\qquad \qquad \qquad \qquad \qquad +(\ell - 1)\ell(\ell+1)(\ell+2)\left| \pert{v}\elem \right|^2   \Bigg).
\end{align}
Here, for convenience, we have defined the degree-$\ell$ and order-$m$ velocity field (per forcing magnitude) as
\begin{align}
    \pert\vel\elem = \pert{u}\elem Y\elem \rhat + \pert{v}\elem \left( \frac{\partial Y\elem}{\partial \theta} \that + \frac{1}{\sin \theta} \frac{\partial Y\elem}{\partial \phi} \phat\right),
\end{align}
where
\begin{equation}
    \pert{u}\elem = \frac{\ell(\ell+1)}{r^2} S\elem \qquad\text{and}\qquad
    \pert{v}\elem = \frac{1}{r} \frac{d S\elem}{dr}
\end{equation}
generate the radial and tangential velocity components (c.f.,~\eqref{eq:velocity-components-S}). With the total dissipation-rate density, we write the dimensionless total dissipation rate of the core as
\begin{equation}
    \label{eq:diss_viscous}
    \dot{E}_\nu(\text{core}) = \int_0^1 \solidangle{e}_\nu r^2 \,\infd r.
\end{equation}

\section{Forcing by solid-mantle motion}
\label{app:solid-mantle-forcing}

In models of a uniform-density core beneath a rigid mantle (i.e., without BMO), the tidal potential cannot directly drive flow in the core. However, the tidal potential can cause flexure of an elastic mantle, and hence motion of the boundary between the core and the mantle.  This motion is typically imposed as a boundary condition on the core; it forces the flow \citep[e.g.,][]{legal2015flows}. However, if the fluid density is non-uniform (e.g., stratified), then the tidal potential can drive a flow directly. In our case, this potential acts on the density stratification of the core and BMO---so far, we have treated the solid mantle as being rigid. However, we anticipate that mantle motion also contributes to flow in the core/BMO system.  Therefore, in this appendix, we compute the inviscid response to an imposed, non-zero tidal motion of the solid mantle, $\vel_S$.

\begin{figure}
    \centering
    \includegraphics[width=\linewidth]{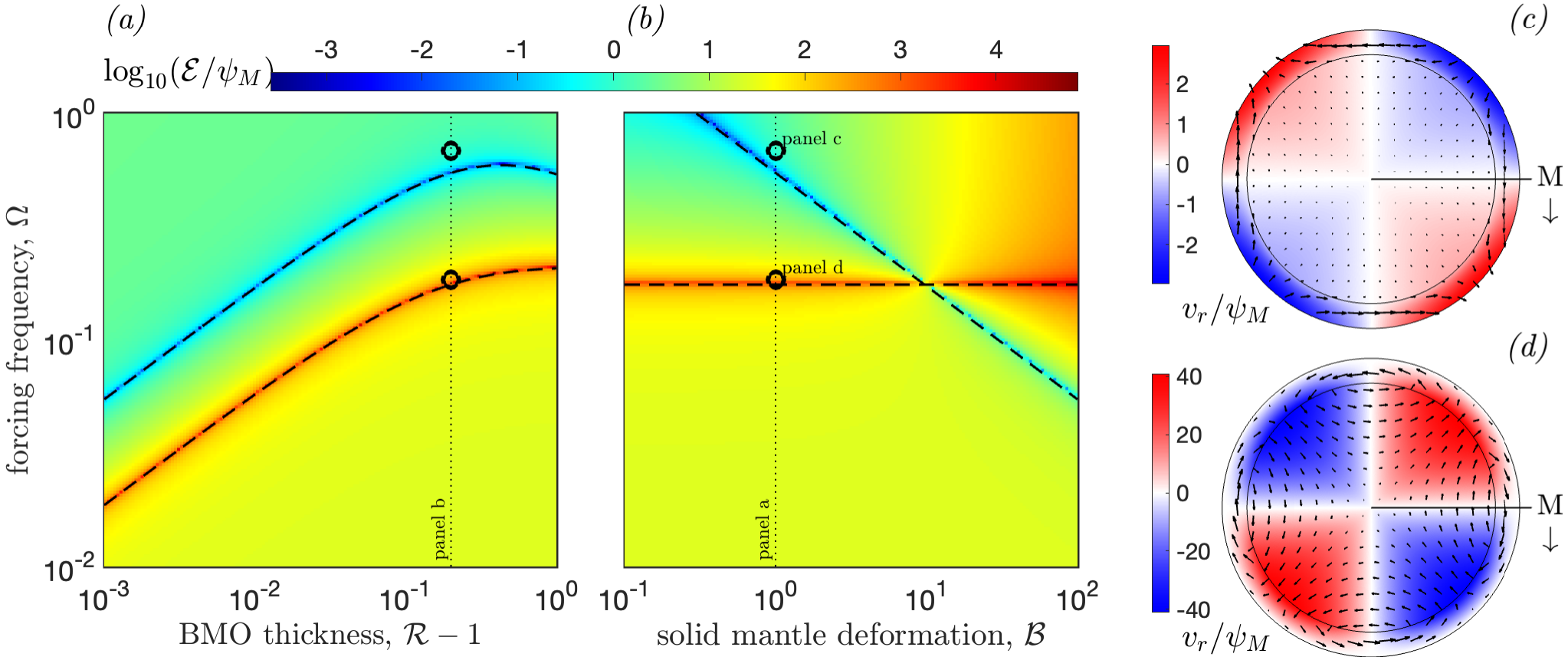}
    \caption{Solutions for non-zero tidal deformation of the solid mantle. \textit{(a)} Ellipticity per forcing magnitude as a function of frequency and BMO thickness with fixed $\curlyb=1$. Circles indicate conditions used in panels~c and d. Dashed lines are inviscid predictions for $\Omega_*$ (eq.~\ref{eq:inviscid-resonance-trend}) and $\Omega_\circ$ (eq.~\ref{eq:inviscid-antiresonance-trend}). \textit{(b)} As in panel~a but with fixed $\curlyr=1.2$. \textit{(c)} Equatorial flow portrait for near-antiresonant case with $\Omega = 0.95\Omega_\circ$. \textit{(d)} Equatorial flow portrait for near-resonant case with $\Omega = 1.05\Omega_*$. }
    \label{fig:mantle-B-appendix}
\end{figure}

Recall the boundary condition~\eqref{eq:bc-dimensional-rO}, which is a no-slip condition at $r=r_O$. This condition is linearised to apply on the sphere $r=R_O$. To obtain an expression for $\vel_S (\curlyr, \theta, \phi, t)$, we prescribe the solid mantle's tidal response to the 2,2 tidal potential in terms of an ellipticity $\beta$. We assume that the solid-mantle ellipsoid has a long axis aligned in the direction of the Moon.  Defining $\pert{r}_O$ as a small, dimensional perturbation to $R_O$ we have ellipticity $\beta \sim 2|\pert{r}_O|/R_O$. From this we can infer that $\pert{r}_O = (\beta R_O/6)Y_2^2(\theta,\phi+\omega t)$. Then we use $\rhat\cdot\vel_S = \partial_t\pert{r}_O$ and non-dimensionalise to obtain
\begin{equation}
    \label{eq:rvel-solid-mantle}
    \rhat\cdot\vel_S(r\sim R_O) = (i/2)\Omega\curlyb\curlyr\tidepotm Y_2^2(\theta,\phi+\Omega t) + \cc, \qquad\text{where }\curlyb \equiv \beta/3\tidepotm.
\end{equation}
 This speed is incorporated into the right-hand side of the boundary-condition system \eqref{eq:boundary-condition-linear-system}.

The non-dimensional ellipticity of the solid mantle can be expressed as  
\begin{equation}
    \curlyb = 6 h_{22} \frac{M_c}{M_E} \left( \frac{R_E}{R_C} \right)^3,
\end{equation}
where $M_E$ and $R_E$ are the mass and radius of Earth, respectively, and $h_{22}$ is the radial-displacement tidal Love number \citep{love1911ProblemsGeodynamics}. The present day Love number at Earth's surface is $h_{22}\sim0.6$ \citep{na2010whole-earth-oscillation}, and the maximum possible value is for a fully fluid mantle, $h_{22} = 5/2$ \citep{love1911ProblemsGeodynamics}. These limits indicate that $5 \lesssim \curlyb \lesssim 30$ throughout the history of Earth. 

The 2,2 tide can also create tangential motion on the base of the solid mantle; this is quantified in terms of the Shida number \citep{shida1912}. However, the tangential component of velocity is $\sim$5$\times$ smaller than the radial component, and it is only relevant if coupled to the flow through the viscosity. Here we consider an inviscid flow in the BMO, and so we need only the radial component of $\vel_S$.

For the inviscid case, previously considered in \S\ref{sec:inviscid-solution}, the solution with non-zero $\curlyb$ uses the coefficients
\begin{subequations}
\label{eq:inviscid-analytical-vS}
\begin{align}
    \label{eq:inviscid-A-vS}
    A\elem &=\frac{(1-\roc)/(\ell+1) + \roc\frac{5}{9}\left(\curlyb\curlyr^6\Omega^2\delta_{\ell 2}\delta_{m 2}\right)/\left(1-\curlyr^5\right)}{\frac{\ell\left(1-\roc\right)}{m^2\Omega^2}\left(\frac{1}{3}-\frac{1-\roc}{2\ell+1}\right)  - \frac{\roc}{1-\curlyr^{-2\ell-1}}\left(\curlyr^{-2\ell-1}+\frac{\ell}{\ell+1}\right)- 1},\\
    \label{eq:inviscid-BCr-vS}
    B\elem&=\frac{1}{1-\curlyr^{2\ell+1}}\left(A\elem - \tfrac{1}{3}\curlyb\curlyr^6\Omega^2\delta_{\ell 2}\delta_{m 2}\right),\\ 
    C\elem &=\frac{1}{1-\curlyr^{-2\ell-1}}\left(A\elem - \tfrac{1}{3}\curlyb\curlyr\Omega^2\delta_{\ell 2}\delta_{m 2}\right),
\end{align}
\end{subequations}
with $r\elem$ unchanged from \eqref{eq:inviscid-analytical-coefs}. Since both $r\elem$ and the denominator of $A\elem$ are unchanged, the resonance frequency $\Omega_*$ of equation~\eqref{eq:inviscid-resonance-trend} is unchanged. The resonance trend is shown in the plot of $\ellipticity/\tidepotm$ as a function of $\Omega$ and $\curlyr-1$ in Figure~\ref{fig:mantle-B-appendix}a. 

Figure~\ref{fig:mantle-B-appendix}a has $\curlyb=1$ and shows that this additional forcing creates an antiresonance, in which the ellipticity $\ellipticity$ of the CMB is zero. The antiresonance occurs at a frequency $\Omega_\circ$ with
\begin{equation}
    \label{eq:inviscid-antiresonance-trend}
    \Omega_\circ^2 = \frac{3}{5}\frac{(1-\roc)(\curlyr^5-1)}{\roc\curlyb\curlyr^6} \qquad\text{(inviscid)}.
\end{equation}
The scaling of $\Omega_\circ$ with $\curlyb^{-1/2}$ means that at some value of $\curlyb$, the resonance and antiresonance trends cross.  The trend in $\Omega_\circ$ and crossing with $\Omega_*$ is evident in Figure~\ref{fig:mantle-B-appendix}.  It is interesting to note that the solution is undefined ($A_2^2=0/0$) at the crossing point.

The phase lag between the tidal ellipsoid of the CMB and the Moon is $\Delta\phi$.  Because the solid-mantle ellipsoid is prescribed to align in the direction of the Moon, $\Delta\phi$ is also the phase lag between the CMB ellipsoid and the solid-mantle ellipsoid, and hence it plays a critical role in establishing the antiresonance.  The antiresonance occurs when mantle-driven flow within the BMO destructively interferes with forcing of the system by the tidal potential.  The case shown in Fig.~\ref{fig:mantle-B-appendix}c is very close to this situation (the small difference from antiresonance means that flow in the core is non-zero, which is helpful in visualising the phase.) 

To clarify the variation in behaviour across resonance and anti-resonance, we conduct the thought-experiment of slowly increasing the forcing frequency $\Omega$ while holding $\curlyb=1$ and all other parameters constant.  Along the way, we pause to examine the ellipticity of the CMB at various points. This corresponds to ascending the dotted line in Fig.~\ref{fig:mantle-B-appendix} panels~(a) and (b).  At $\Omega\approx10^{-2}$, the $\ellipticity$ is essentially at its equilibrium value $\equil{\ellipticity}$, meaning that the dominant physical balance is between the tidal potential and the self potential---the tide is a quasi-static displacement of the CMB with $\Delta\phi=0$. In other words, on the line connecting the centres of Earth and the Moon, the acceleration is zero. Flow driven by motion of the solid mantle is negligible.  This situation holds as $\Omega$ increases until it approaches $\Omega_*$, the resonant frequency associated with the 2,2 eigenmode.  For $\Omega\le\Omega_*$, the CMB ellipsoid is aligned with the Moon. 

As $\Omega$ exceeds $\Omega_*$, the physical balance and $\Delta\phi$ change discontinuously.  The lag switches to $\pi/2$ (long axis of the equatorial-plane ellipse is perpendicular to the Earth--Moon line) and the acceleration is now a maximum directly beneath the Moon. This means that the physical balance is between the inertial force and the tidal potential.  A near-resonant solution in this regime is shown in Fig.~\ref{fig:mantle-B-appendix}d. The forcing by the solid mantle remains negligible.

With increasing frequencies above $\Omega_*$, the ellipticity of the CMB decreases because accelerations are reversed before their full effect of the velocity (and displacement) is accumulated.  With diminishing tidal flow, the role of solid-mantle forcing becomes appreciable.  However in this regime, the solid mantle forcing has a phase difference of $\pi/2$ to the ellipticity: its downward flow collides with the core's upward flow.  As $\Omega$ increases toward $\Omega_\circ$, the a destructive interference between the tidal potential and the prescribed mantle takes hold.  This is illustrated in Fig.~\ref{fig:mantle-B-appendix}c.

When $\Omega=\Omega_\circ$ the system is at antiresonance, with full cancellation of the tide in the core and hence zero velocity there.  The flow in the BMO remains strong (as required to create the antiresonance).  At higher frequencies, where $\Omega$ exceeds $\Omega_\circ$, the potency of the tidal potential continues to decline, but ellipticity increases again.  This increase is driven by solid-mantle forcing; the response lag $\Delta\phi$ with respect to the forcing is zero. The ellipticity reaches a plateau as $\Omega\to\infty$.  

A more rigorous approach would be to assume a viscoelastic model for the mantle and solve force-balance equations for its dynamics. These dynamics would be coupled to the core and BMO via the boundary conditions at $r_O$.  This would enable self-consistent prediction of the phase lag and ellipticity at each of the subdomain boundaries.  While some features of the solution may be conserved, the high-frequency behaviour would certainly differ.  While such a model is appealing, it would be considerably more complex to solve and interpret; it is left for future work.  

\section{Parameterised orbital--BMO trajectory}
\label{app:trajectory}

\cite{farhat2022resonant} provide a model of Earth--Moon orbital distance $\Rorb$ and Earth length of day (LOD, $T$) as a function of time since the Moon-forming impact at about 4.5~Ga. We approximate their models with the simple, smooth parameterisations
\begin{subequations}
    \label{eq:orbital-evolution-model}
    \begin{align}
        \Rorb(t) &= R_{\text{orb}, i} + (R_{\text{orb}, f} - R_{\text{orb}, i})\,(t/4.5~\text{Ga})^{1/3},\\
        T(t) &= T_i + (T_f-T_i)\,(t/4.5~\text{Ga})^{2/3},
    \end{align}
\end{subequations}
where subscripts $i,f$ indicate the initial and final values, respectively (see Table~\ref{tab:orbital-evolution}) and $t$ is the elapsed time since the Moon-forming impact in billions of years.

\begin{table}
    \centering
    \begin{tabular}{ccll}
        Symbol & value & units & note/reference \\\hline
        $R_{\text{orb}, i}$ & $9.6\times10^4$ & km & $\sim15 R_E$, \cite{farhat2022resonant} \\
        $R_{\text{orb}, f}$ & $3.85\times10^5$ & km & $\sim60 R_E$ \\
        $T_{\text{orb}, i}$ & 6 & hr & \cite{farhat2022resonant} \\
        $T_{\text{orb}, f}$ & 24 & hr &  \\
        $H_{O i}$ & 100--400 & km & \cite{schaeffer2025energetically} \\
        $\tau$ & 1.28, 0.85 and 0.92 & Ga & \cite{labrosse2007crystallizing}
    \end{tabular}
    \caption{Parameter values relevant to the orbital--BMO evolution model. The decay constants of 1.28, 0.85 and 0.92~Ga correspond to initial BMO thicknesses of 100, 200 and 400~km, respectively.}
    \label{tab:orbital-evolution}
\end{table}

\cite{schaeffer2025energetically} provide a model of the evolution of the BMO thickness $H_O(t)$, from which we parameterise $R_O$, the outer radius of the BMO,
\begin{equation}
    \label{eq:orbital-evolution-model-BMO}
    R_O(t) = R_C + H_{Oi}\,\e^{-t/\tau},
\end{equation}
where $\tau$ is a decay constant.  This exponential was derived by \cite{labrosse2007crystallizing} as an approximation of a more detailed theory.  The range of initial BMO thickness given by \cite{schaeffer2025energetically} spans 100 to $\sim$400~km. The full set of parameter values used to compute the orbital--BMO trajectory is given in Table~\ref{tab:orbital-evolution}.

\bibliographystyle{jfm}
\bibliography{references}

\end{document}

%% file: newcommands.tex
\newcommand{\infd}{\text{d}}
\newcommand{\Grad}{\boldsymbol{\nabla}}
\newcommand{\Div}{\Grad\cdot}
\newcommand{\Curl}{\Grad\times}
\newcommand{\delsq}{\nabla^2}
\newcommand{\pd}[2]{\frac{\partial{#1}}{\partial{#2}}}

\newcommand{\fd}[2]{\frac{\text{d}{#1}}{\text{d}{#2}}}
\newcommand{\jump}[1]{\left\llbracket{#1}\right\rrbracket}
\newcommand{\e}{\text{e}}
\newcommand{\dirac}{\delta}
\newcommand{\identity}{\boldsymbol{I}}
\newcommand{\pos}{\boldsymbol{x}}
\newcommand{\rpos}{\boldsymbol{r}}
\newcommand{\vel}{\boldsymbol{v}}
\newcommand{\nvec}{\mathbf{n}}
\newcommand{\tvec}{\mathbf{t}}

\newcommand{\zhat}{\widehat{\boldsymbol{z}}}
\newcommand{\rhat}{\widehat{\boldsymbol{r}}}
\newcommand{\that}{\widehat{\boldsymbol{\theta}}}
\newcommand{\phat}{\widehat{\boldsymbol{\phi}}}
\newcommand{\stressten}{\boldsymbol{\sigma}}
\newcommand{\devstressten}{\boldsymbol{\tau}}
\newcommand{\strainrten}{\dot{\boldsymbol{\epsilon}}}

\newcommand{\legendre}[2]{\mathcal{P}_{#1}^{#2}}
\newcommand{\sph}[2]{Y_{#1}^{#2}}
\newcommand{\elem}{^m_\ell}

\newcommand{\lmsumh}{\sum_{\ell=2}^{\infty}\sum_{m=1}^{\ell}}
\newcommand{\pert}[1]{\Tilde{#1}}
\newcommand{\equil}[1]{\Breve{#1}}
\newcommand{\solidangle}[1]{\overset{\frown}{#1}}
\newcommand{\re}{\text{Re}}
\newcommand{\reo}{\re_O}
\newcommand{\rem}{\text{Rm}}
\newcommand{\pr}{\text{Pm}}
\newcommand{\Ei}{\text{Ei}}
\newcommand{\Ek}{\text{Ek}} 
\newcommand{\Rorb}{R_\text{orb}}
\newcommand{\curlyr}{\mathcal{R}}
\newcommand{\curlyb}{\mathcal{B}}
\newcommand{\ellipticity}{\mathcal{E}}
\newcommand{\tidepot}{\psi}
\newcommand{\tidepotm}{\psi_M}
\newcommand{\Uel}{U_\ellipticity}
\newcommand{\roc}{\rho_{_{OC}}}
\newcommand{\noc}{\nu_{_{OC}}}
\newcommand{\cc}{\mathrm{c.c.}}